\documentclass[aps,twocolumn,showpacs,superscriptaddress]{revtex4}
\usepackage{epsfig}
\usepackage{times}

\begin{document}

\title{Cooperation driven by success-driven group formation}

\author{Attila Szolnoki}
\email{szolnoki@mfa.kfki.hu}
\affiliation{Institute of Technical Physics and Materials Science, Centre for Energy Research, Hungarian Academy of Sciences, P.O. Box 49, H-1525 Budapest, Hungary}

\author{Xiaojie Chen}
\email{xiaojiechen@uestc.edu.cn}
\affiliation{School of Mathematical Sciences, University of Electronic Science and Technology of China, Chengdu 611731, China}

\begin{abstract}
In the traditional setup of public goods game all players are involved in every available groups and the mutual benefit is shared among competing cooperator and defector strategies. But in real life situations the group formation of players could be more sophisticated because not all players are attractive enough for others to participate in a joint venture. What if only those players can initiate a group formation and establish a game who are successful enough to the neighbors? To elaborate this idea we employ a modified protocol and demonstrate that a carefully chosen threshold to establish joint venture could efficiently improve the cooperation level even if the synergy factor would suggest a full defector state otherwise. The microscopic mechanism which is responsible for this effect is based on the asymmetric consequences of competing strategies: while the success of a cooperator provides a long-time well-being for the neighborhood, the temporary advantage of defection cannot be maintained if the protocol is based on the success of leaders.
\end{abstract}

\maketitle

\section{Introduction}
Social dilemmas are frequently captured within the framework of public goods game where two basic strategies compete: while cooperators contribute to the common pool, defectors do not, but only enjoy the benefit of mutual efforts \cite{sigmund_10}. Not surprisingly, if a player's goal is to reach a better individual position by following a more successful strategy then the system can easily be trapped into the so-called ``tragedy of the common state" where full defection forces a global minimum for all \cite{hardin_g_s68,gibbons_92}. Nevertheless, it is our everyday life experience that cooperation embraces us \cite{ledyard_97}, which is a clear indication that a fundamental detail is ignored by our basic approach.

Several adequate suggestions have been made in the last decade which try to address the conflict of experiment and theory \cite{nowak_11,sicardi_jtb09,fu_srep12,rezaei_pa12,pena_pre09,arenas_jtb11,mobilia_pre12,wu_t_pre09,requejo_pre12b,sasaki_prsb07,liu_yk_epl13,vilone_jsm11,hindersin_pcbi15,suzuki_r_pre08,mobilia_csf13,javarone_epjb16,jordan_pnas16}. One of these research directions assumed that more sophisticated strategies should be used, which go beyond the simplest unconditional cooperator and defector behaviors \cite{press_pnas12,burton-chellew_pnas16}. In case of conditional cooperation, for instance, we may assume that some players cooperate only if a certain portion of group members are willing to cooperate \cite{szolnoki_pre12,chen_xj_pone12b}. Or, from a different viewpoint, some players are tolerant toward a minimal level of defection and change attitude only if the number of defectors exceeds a threshold level \cite{szolnoki_pre15,szolnoki_njp16}. But we can also quote the simplest approach which applies a similar unconditional strategy, loner, who prefers not to participate in the joint venture, but choose a moderate fixed income \cite{hauert_s02,semmann_n03}.

A further conceptually different research avenue warrants that players are not randomly mixed, as it is supposed in the early works, but instead we should consider a sort of interaction graph which determines the possible partners of every individual \cite{nowak_n92b}. The most important consequences of this concept are limited and permanent interactions which establish the chance for the so-called network reciprocity mechanism to work \cite{santos_prl05,szabo_pr07}. The latter is considered as one of the fundamental ways how to escape the trap raised in the introductory paragraph \cite{nowak_s06}. Let us stress that there are several other interesting possibilities, such as the heterogeneity of players \cite{szolnoki_epl07,perc_pre08b}, or special character of interaction topology \cite{santos_prl05,wang_z_epl12,gomez-gardenes_srep12}, which could also be helpful to support cooperator strategy. For further details we refer a recent review of this game where comprehensive overview can be found about our current understanding \cite{perc_jrsi13}.

We should note that most of the mentioned works assume a certain symmetry on how the interaction graph is used. More precisely, a focal player, who organizes a game with actual neighbors,  also participates in the games organized by nearest neighbors. In this way a player is always involved in several independent games which are determined by the interaction graph. This simple assumption, however, ignores a significant real-life experience. Namely, to participate in a joint venture is not always attractive to every potential group members. Consequently, players are reluctant to join if an unsuccessful player organizes a game but they are more enthusiastic when a successful player, having high payoff, is announcing a game. The simplest way to consider this attitude is if we introduce a threshold level of payoff which should be reached by all players who want to organize a game. Those who fail to fulfill this criterion cannot establish a group hence cannot organize a game, but can only participate in the games of others, which remains the only way to collect some payoff for them.

It is important to emphasize that the proposed protocol is ``strategy-neutral", i.e., there is no direct support of cooperator strategy. When the concept of reputation is used, for instance, then the latter criterion is not fulfilled because the applied dynamics gives a direct advantage for those who cooperate and behave positively toward others \cite{milinski_n02}. Interestingly, similar claim can also be raised when punishment (of defectors) or reward (of cooperators) are used because in these cases we always assume a sort of cognitive skill from players to identify others' strategies \cite{rezaei_ei09, chen_q_csf16, cong_r_epl16}. In our present model we do not assume such an extra ability about players, but only suppose that they recognize the resulting payoff of their partners. Since both cooperator and defector strategy may gain individual success, the consequence of the proposed protocol is far from trivial. In the next section we describe the employed evolutionary game, whereas Section~\ref{results} is devoted to the presentation of our observations. Finally in the last section we summarize and discuss their implications.

\section{Model definition}

The public goods game is staged on a square lattice where players can cooperate or defect. According to the applied interaction topology every player may participate in $N = 5$ different games which is organized by the focal individual of a von Neuman neighborhood and four nearest neighbors. As a consequence, the individual payoff of player $i$ is calculated as $\Pi_i = \sum_j \Pi_i^j$, where $\Pi_i^j$ is the income of $i$ from the game which is organized by the neighboring player $j$. It is a fundamental point, however, that a player $j$ can organize a game only if $\Pi_j$ collected in the previous round exceeds a critical threshold $H$. If it is fulfilled then all neighboring cooperator players contribute $c=1$ to the public good which is multiplied by a synergy factor $r$. After this the enhanced amount is shared equally among all group members no matter if they contributed or not. In other words, the success in collecting high payoff in the last round gives a merit to distinguished players to organize a game in the subsequent round. In this way the status of all players can be characterized by a four-state model. In particular, $C_l$ denotes the status of those cooperators who were unsuccessful in the last round therefore their merit is ``low" hence they can only contribute to the games organized by their neighbors. $C_h$ denotes ``high-merit" cooperators who were successful enough to collect the requested payoff in the last round therefore they can also organize their own game. Similarly, unsuccessful defectors are denoted by $D_l$ whose only chance to collect payoff is to participate in the neighbors' games. Finally, $D_h$ marks the high-merit defectors who are able to organize their own games which offers them an extra way to exploit their neighbors.

Initially each player on site $x$ is designated either as a cooperator or as a defector, and meanwhile is endowed with ``low" or ``high" merit with equal probability. According to an elementary Monte Carlo step a player $x$ is selected randomly whose $\Pi_x$ payoff is calculated in the way we described in the previous paragraph. Let us stress again that the resulting $\Pi_x$ payoff depends not only on the strategies of neighbors but also on their merit values. Next, one of the four nearest neighbors of player $x$ is chosen randomly, and its location is denoted by $y$. Player $y$ also acquires its payoff $\Pi_y$ as previously described for player $x$. Finally, player $x$ imitates the strategy of player $y$ with the probability 
\begin{equation}
s(\Pi_x, \Pi_y) = \frac{1}{{1 + \exp [({\Pi_x} - {\Pi_y})/K]}},
\label{strategy}
\end{equation}
where $K$ determines the level of uncertainty by strategy adoptions \cite{szabo_pre98}. Without the loss of generality we set $K = 0.5$, implying that better performing players are almost imitated, but it is not impossible to adopt the strategy of a player performing worse. Each full Monte Carlo step ($MCS$) involves that all players having a chance to adopt a strategy from one of their neighbors once on average. Beside strategy adoption the merit of players are also updated. Similarly to strategy update here we also considered the possibility of error in judgment or player's deception. Accordingly the merit of a player $x$ is set to be ``high" with probability 
\begin{equation}
m(\Pi_x) = \frac{1}{{1 + \exp [(H - {\Pi_x})/K]}},
\label{merit}
\end{equation}
otherwise its merit is selected to be ``low". For simplicity we used the same level of uncertainty of merit selection, but the extension toward more general model with different noise levels is possible. Note that the stochastic character of merit's update has a similar fundamental role as it has for strategy update. Namely, it allows the system to avoid a frozen trapped state hence revealing the leading cooperator supporting mechanism.  

Depending on the applied threshold value $H$ and the synergy factor $r$ the linear system size was varied from $L = 400 - 6000$ in order to avoid finite size effects. The necessary time to reach the stationary state is varied from $5 \cdot 10^3$ to $10^5$ $MCS$s. Evidently, the evolution is halted if one of the competing strategies extinct or all high-merit players die out. In the latter case, when only low-merit players are present, groups are not formed hence there is no actual interaction between players and the evolution between defectors and cooperators becomes a neutral drift similar to the case of voter-model dynamics \cite{cox_ap86,dornic_prl01}. We have repeated our independent runs 10-1000 times to reach the requested accuracy of presented data.

\section{Results}
\label{results}

\begin{figure}
\centerline{\epsfig{file=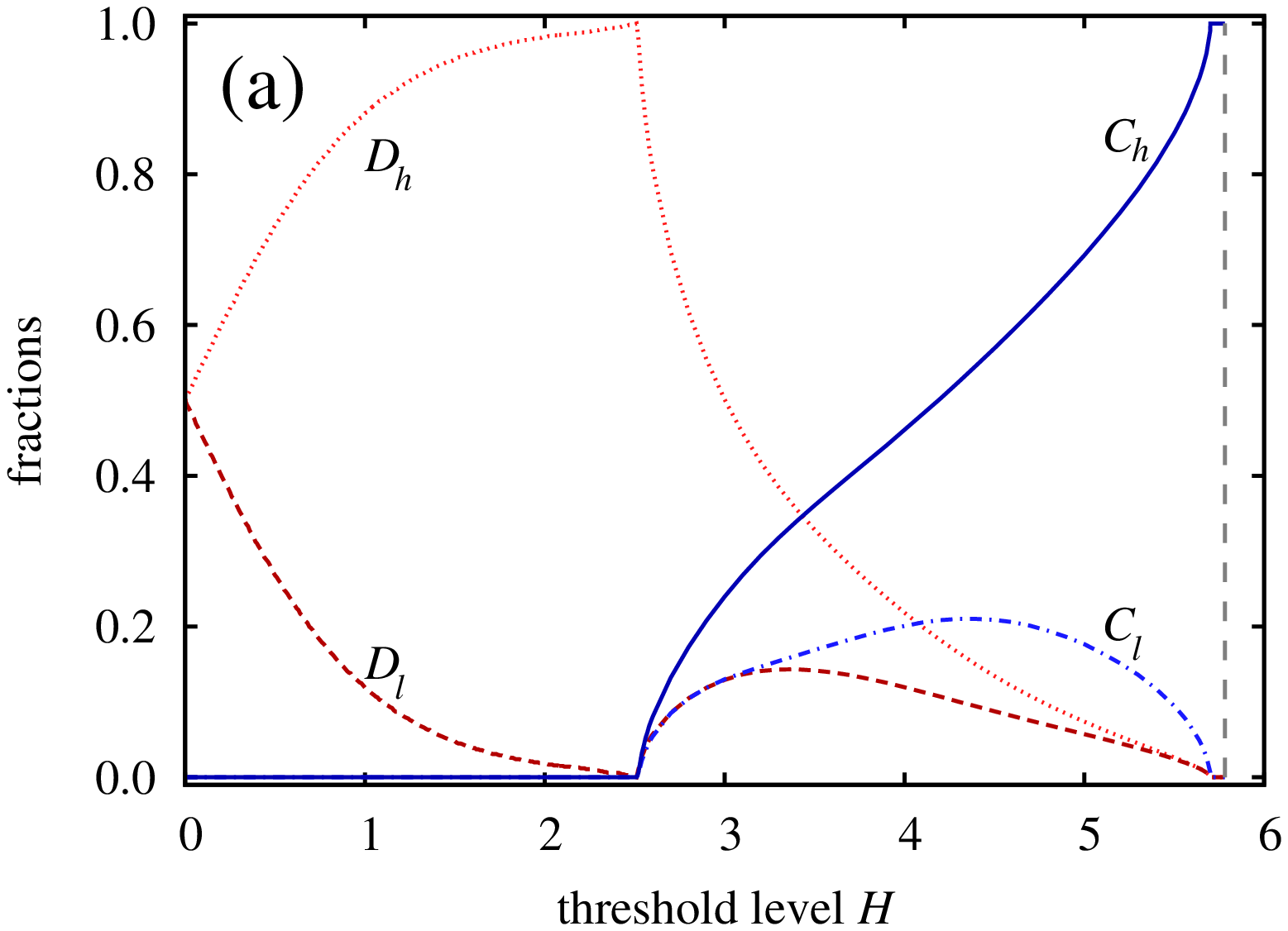,width=8.0cm}}
\centerline{\epsfig{file=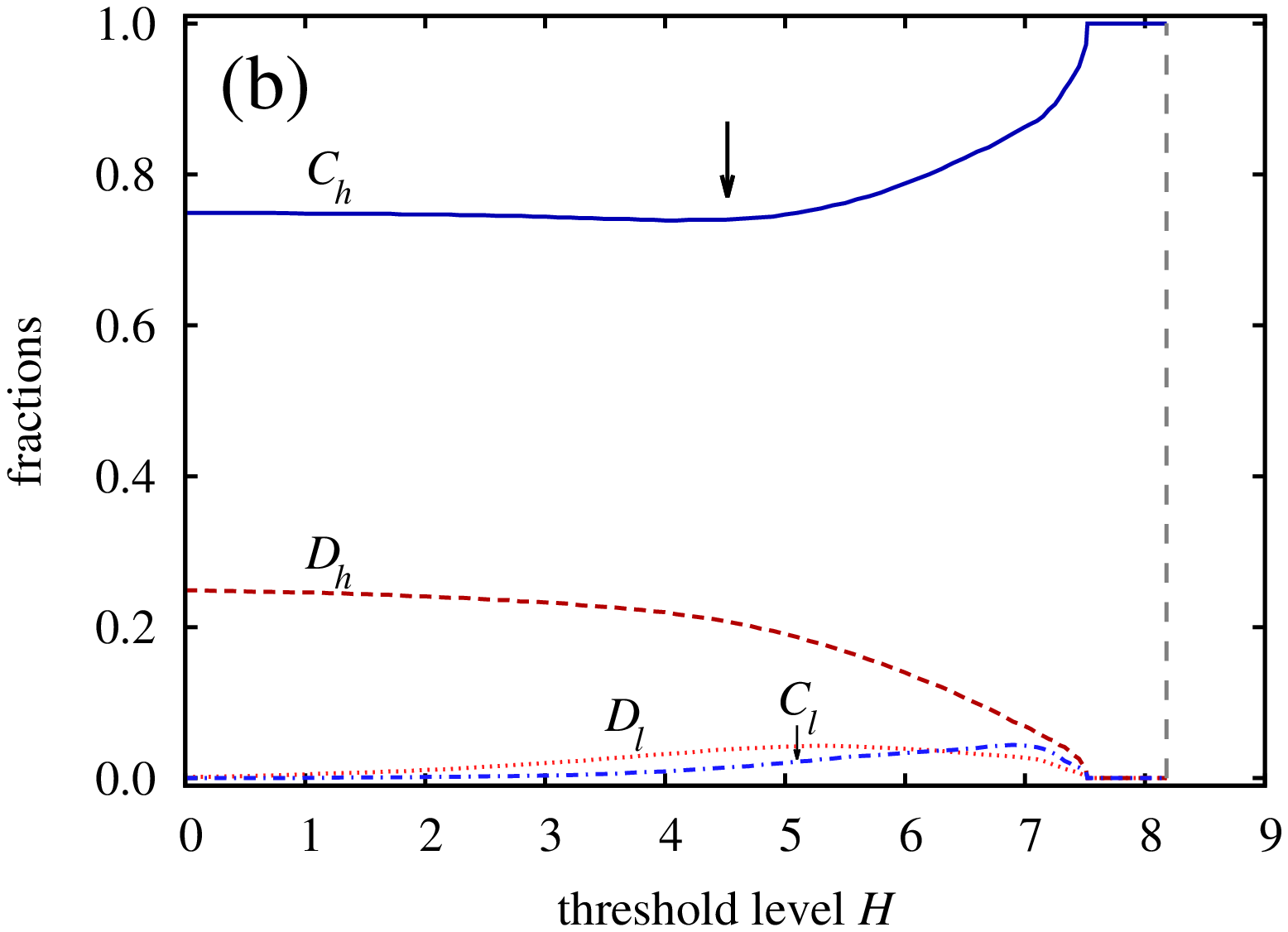,width=8.0cm}}
\caption{(Color online) Stationary fractions of the four possible states of players in dependence on the threshold value $H$ as obtained for $r=3.5$ (in Fig.1(a)) and for $r=4.5$ (in Fig.1(b)). Cooperators (defectors) with high and low merit are denoted by $C_h$ and $C_l$ ($D_h$ and $D_l$) respectively. The application of low $H$ values has no particular impact on the strategy evolution: only defectors survive at low $r$ value, while cooperators and defectors coexist at high $r$ value. If we increase the threshold value then cooperators become dominant gradually and the system reaches the full-cooperator state. Increasing the $H$ value further the system is trapped into a frozen state where both high-merit strategies extinct at an early stage of evolution. The border of this critical $H$ value is marked by thin dashed-line in both panels. In the bottom panel arrow marks the critical $H$ value where success-driven group formation already improves the cooperation level. The error bars are comparable to the width of curves.}
\label{H}
\end{figure}

First we illustrate how the application of success-driven group formation influences the cooperation level at a fixed value of synergy factor. In Figure~\ref{H} we have chosen two representative $r$ values. In the top panel we use $r=3.5$ which would result in a full defector state in the traditional model where all possible groups are formed \cite{szolnoki_pre09c}. As the top panel shows, applying small threshold value has no particular role on the cooperation level: the system still evolves into a full defector state, where the only consequence is the relative fraction of $D_h$ and $D_l$ players characterizing the state when the last cooperator dies out. While $D_h$ and $D_l$ states are represented almost with equal weights for very small $H$ values, the fraction of $D_h$ increases monotonously as we increase the threshold level. When $H$ exceeds the critical $H_c=2.51$ value then a qualitatively new solution emerges during the evolution in which cooperators and defectors coexist. Here the application of success-driven group formation has a positive impact on the cooperation level because significantly high threshold value can only be reached in the vicinity of cooperator players. This selection mechanism could be so powerful that the fraction of cooperators increases gradually by increasing $H$ and the system can evolve to a full cooperator state if $H>5.701$. Interestingly, the full-cooperator state is bordered by a so-called ``frozen" state if the threshold level is too high. In the latter case all initially high-merit players die out soon because they unable to maintain the requested high payoff. As a result, only low-merit cooperators and low-merit defectors remain who are actually unable to collect payoff because there are no properly organized games. In this case Eq.~\ref{strategy} dictates a random strategy update which is conceptually similar to a neutral drift of voter-model \cite{cox_ap86}. Here both full $C$ and full $D$ states are possible destinations via a logarithmically slow coarsening \cite{dornic_prl01} where the probability to terminate to the mentioned states depends on the initial fractions of strategies.

If $r$ is high enough, shown in Fig.~\ref{H}(b), cooperators coexist with defectors in the traditional model due to the well-known network reciprocity mechanism \cite{szolnoki_pre09c}. Similarly to the previously discussed case low $H$ values have no particular role on the stationary state: if $H<H_c=4.52$ for the mentioned $r$ value then the sum of cooperator players equals to the cooperation level that can be obtained in the traditional model where all possible group formations are considered. Above this critical value, marked by an arrow in Fig.~\ref{H}(b), the application of success-driven group forming protocol helps to support cooperator strategy. The full-cooperator state can also be reached if the applied $H$ exceeds the $H=7.52$ critical value. This  phase is bordered by the earlier mentioned frozen region where it is more likely that a finite-size system evolves to a low-merit state than to reach the defector-free destination.

\begin{figure}
\centerline{\epsfig{file=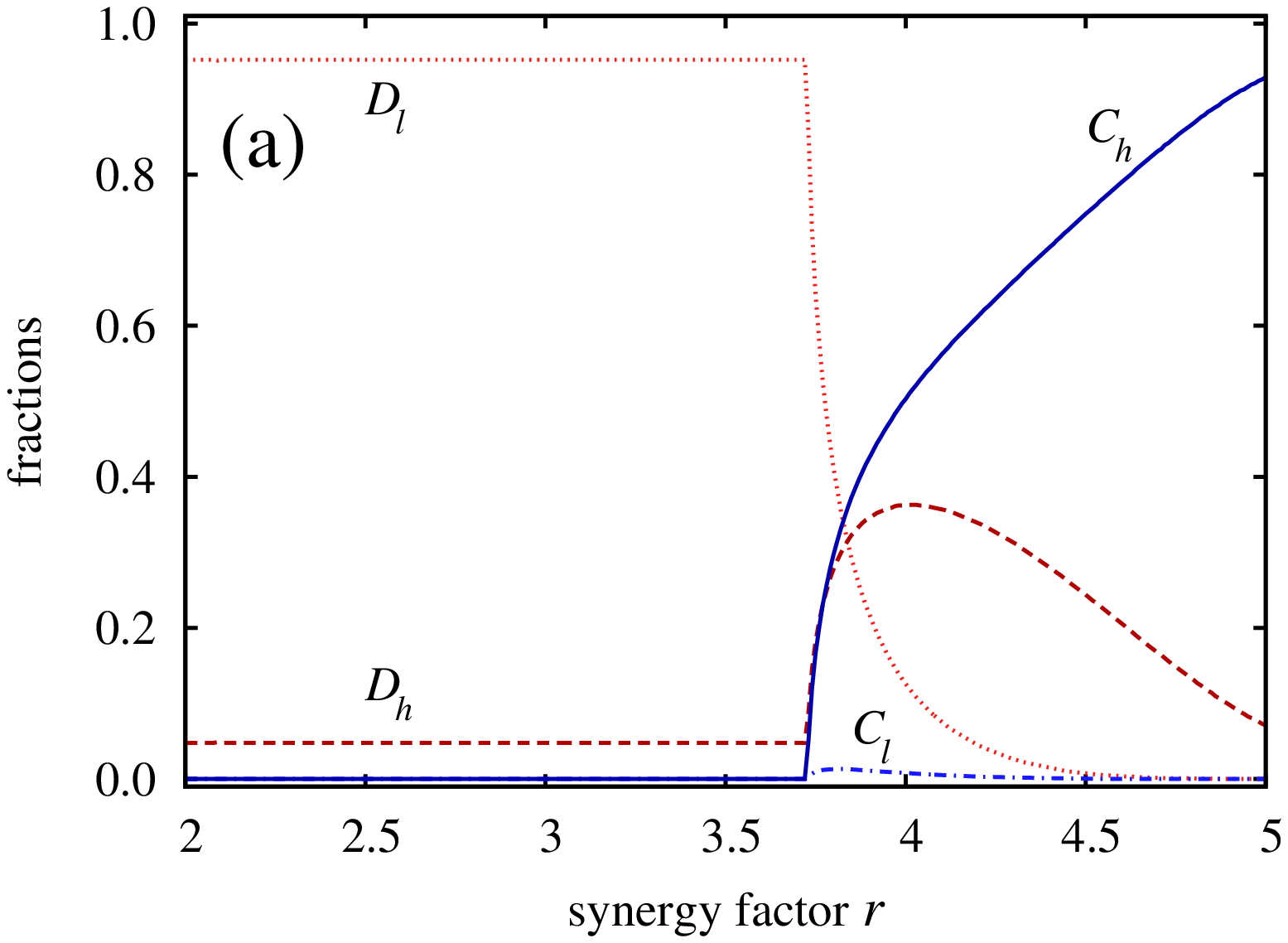,width=8.0cm}}
\centerline{\epsfig{file=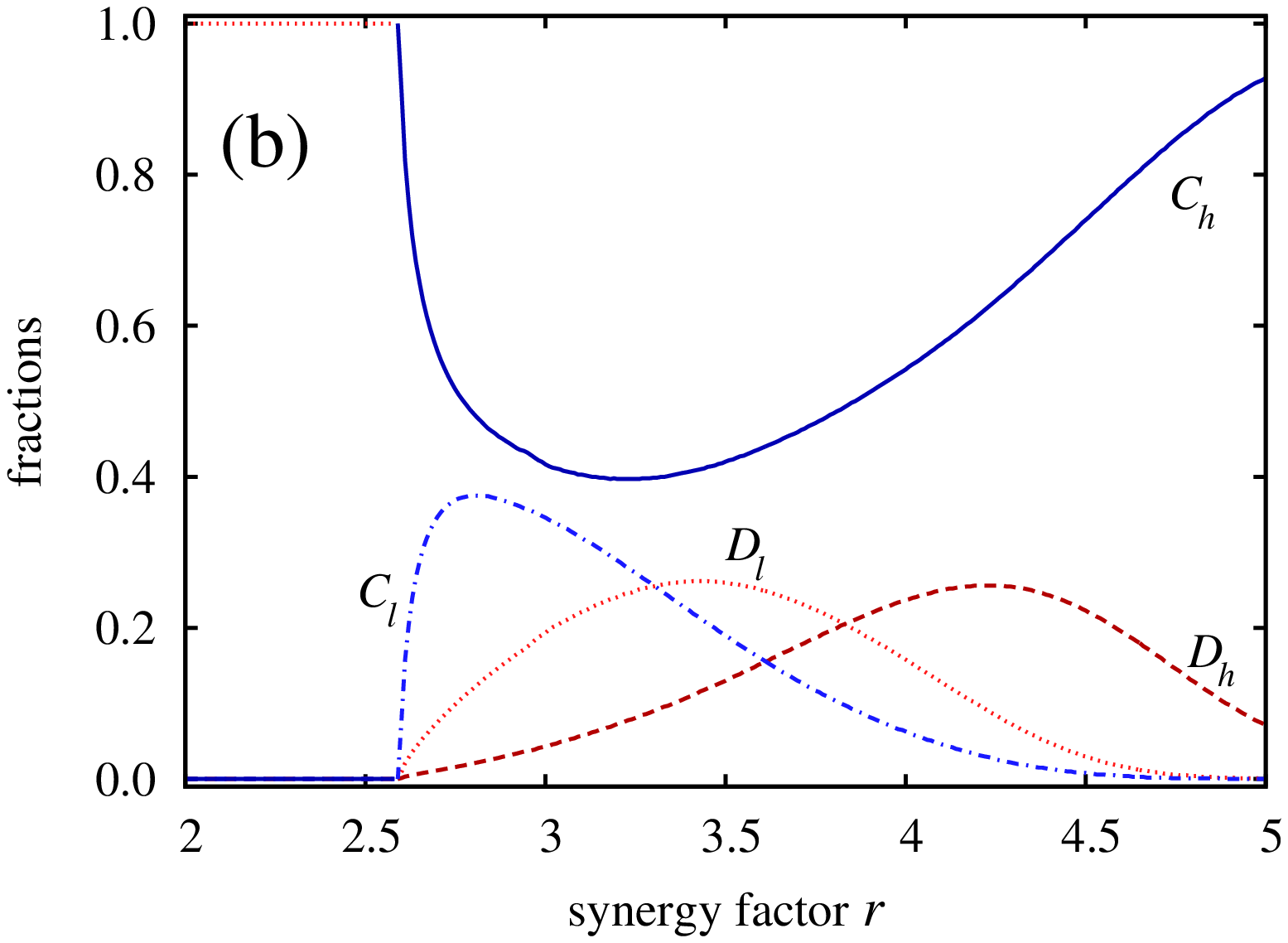,width=8.0cm}}
\caption{(Color online) Stationary fractions of the four possible states of players in dependence on synergy factor $r$ for low ($H=1.5$ in Fig.2(a)) and high threshold value ($H=3.8$ in Fig.2(b)). For low $H$ value the system behavior is similar to the classical model where higher $r$ always results in higher fraction of cooperators. Interestingly, increasing $r$ does not necessarily elevate the cooperation level for high $H$.}
\label{r}
\end{figure}

\begin{figure*}
\centerline{\epsfig{file=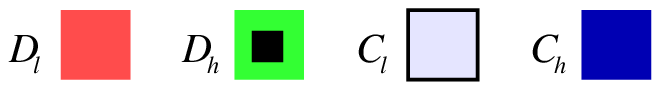,width=5.0cm}}
\centerline{\epsfig{file=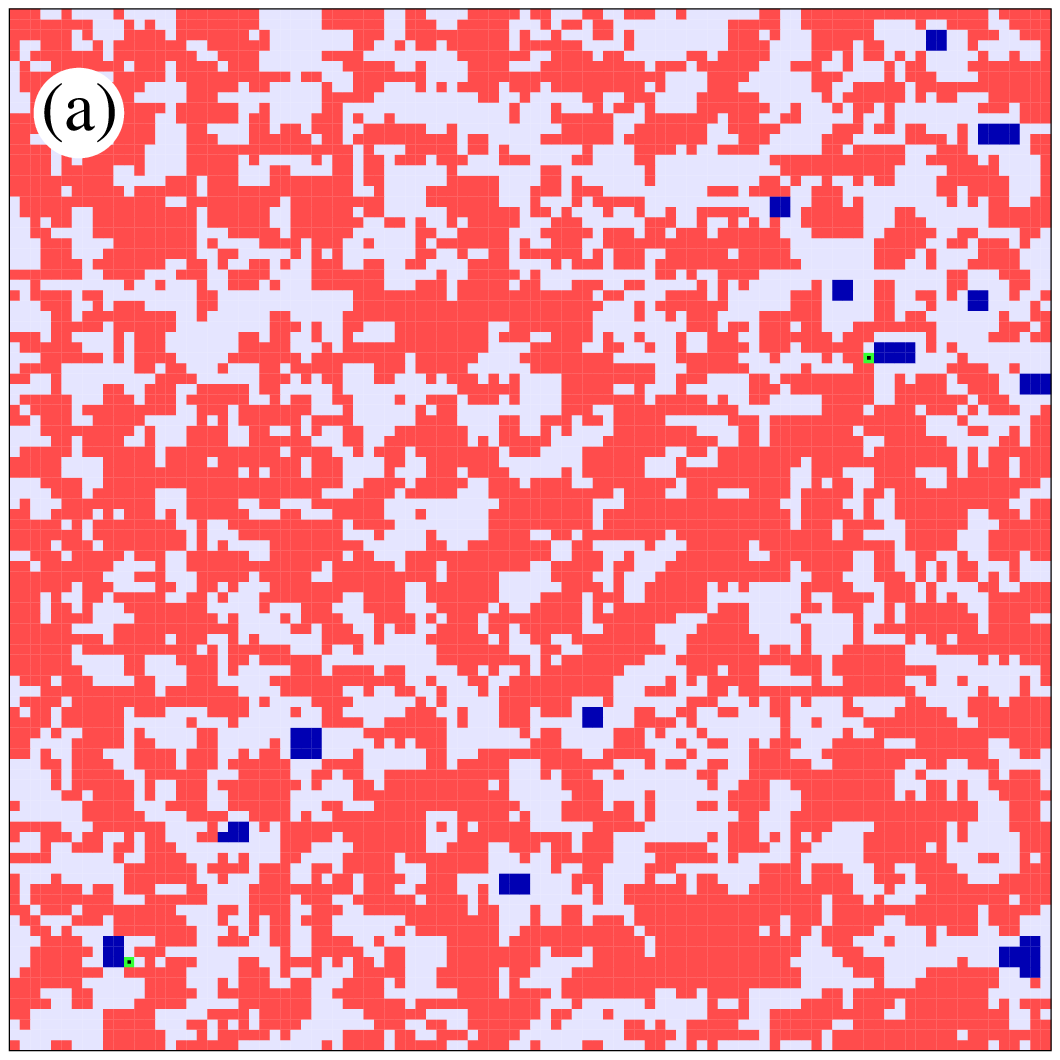,width=4.5cm}\epsfig{file=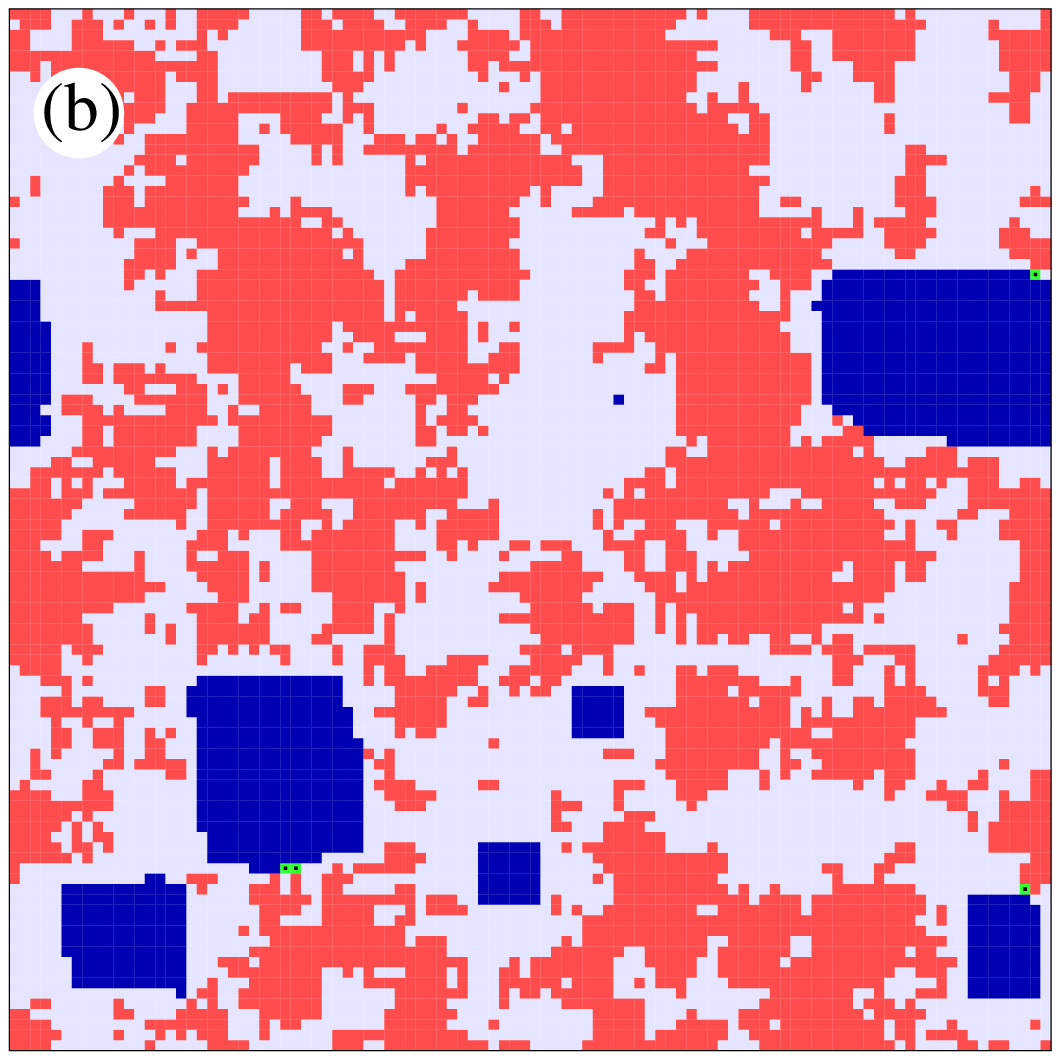,width=4.5cm}\epsfig{file=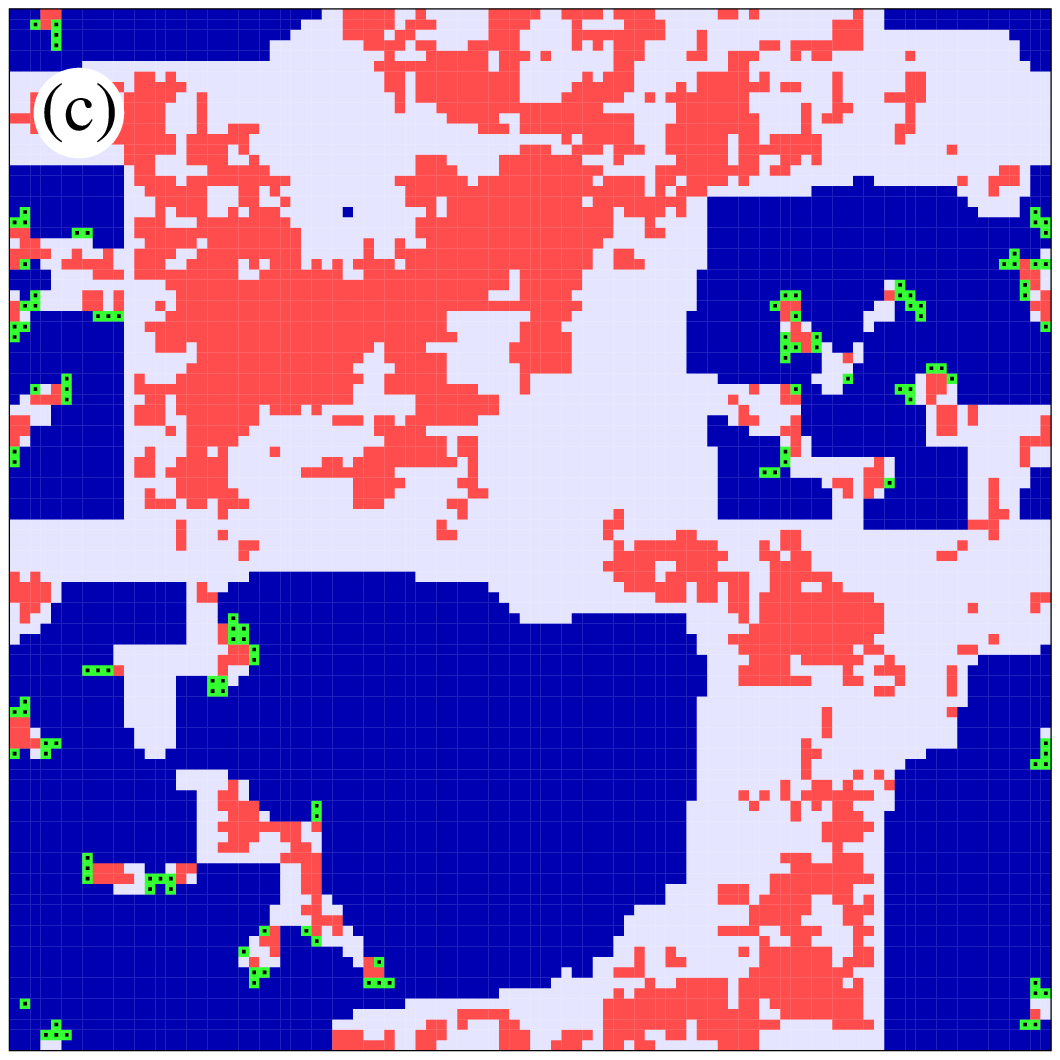,width=4.5cm}\epsfig{file=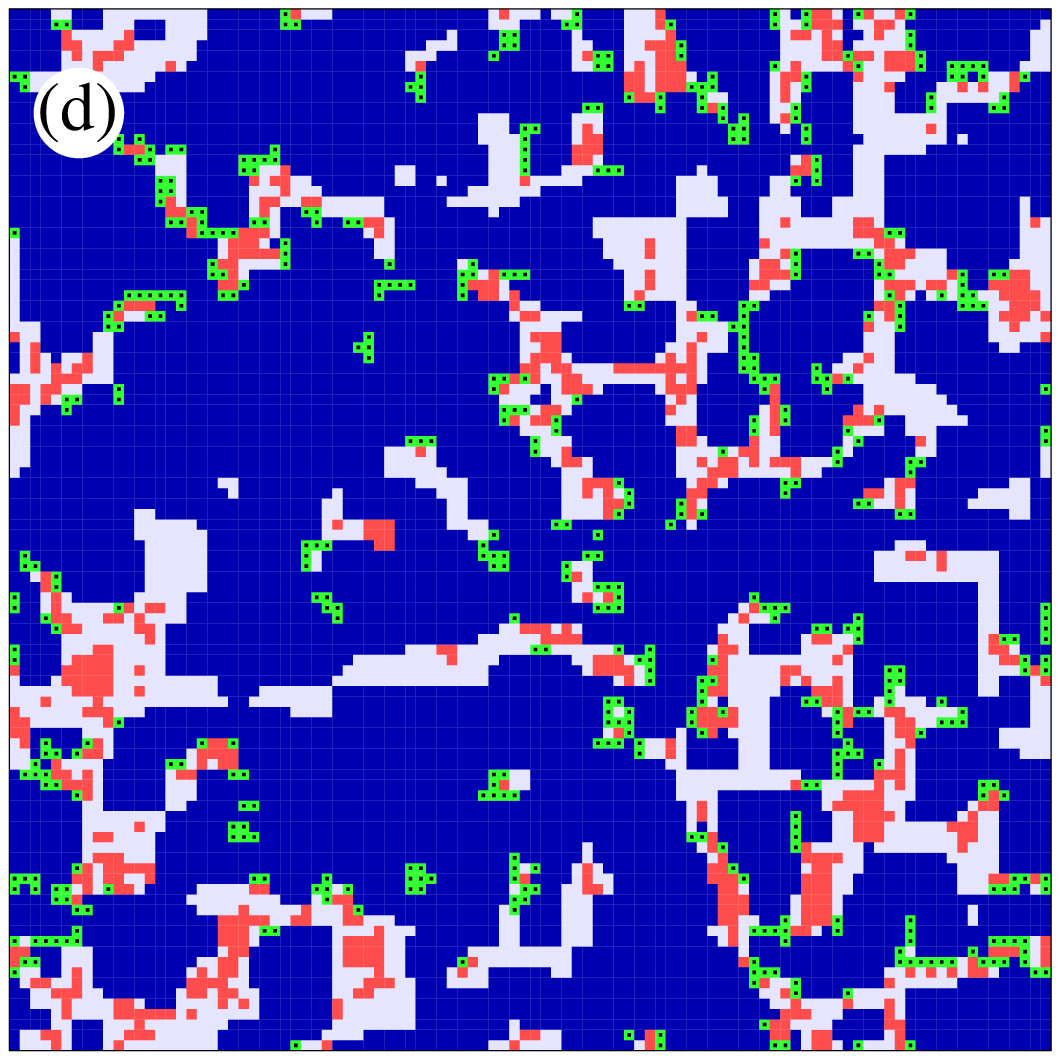,width=4.5cm}}
\caption{\label{snapshot} (Color online) Time evolution of random initial state as obtained for $r=3.5$ and $H=5$ on a square lattice using $L=100$ system size. Cooperators with high merit are marked by dark blue (dark grey), whereas the same strategy with low merit is denoted by light blue (light grey). Defectors with low merit are marked by red (middle grey) and defectors with high merit are denoted by a green color (dotted lighted grey), as indicated by the legend on the top. Snapshots were taken at 10, 200, 400, and 800 $MCS$s. Because of the high threshold value almost all players become in a low-merit state, shown in panel~(a), and only a tiny seed of cooperator domains is capable to collect the necessary high payoff to enjoy high-merit state. Due to a neutral drift between low-merit strategies a slow coarsening starts, as illustrated in panel~(b). Importantly, high-merit cooperator domains grow undisturbed way deep in low-merit domains which offers a protecting shield against defectors. Even if a defector becomes a neighbor of a high-merit cooperator domain, as shown in panel~(c), the resulting high payoff and the related high merit offer just a temporary advantage to $D_h$ player: when a neighboring cooperator follows the victor then the latter's payoff, hence merit, falls down and the mentioned defector will be unable to utilize cooperator neighbors efficiently anymore. In the stationary state, which is shown in panel~(d), the majority of players remain in a cooperator status and just a small portion of players could enjoy temporarily the benefit of defector strategy. Note that the system would be in full-defector state in the absence of success-driven group formation protocol for this low value of synergy factor. Further details are given in the main text.}
\end{figure*}

Our next figure illustrates the impact of increasing synergy factor $r$ when the threshold value is fixed. The top panel of Fig.~\ref{r} shows a representative plot for a low threshold level. Here the application of success-driven group formation protocol has no particular influence on the cooperation level. Namely, if the synergy factor is too low then an initially random system always terminates into a full defector state leaving a mixture of $D_l$ and $D_h$ players. One may claim that the synergy factor is too small for defectors to keep the high-merit state hence $D_h$ players should extinct, too. Indeed, they can only collect high payoff and gain high merit only at the early state of evolution when cooperators are present. When the last cooperator dies out the simulation is halted, hence the fraction of $D_h$ players reflects this stage of the evolution. By increasing $r$ further the cooperation level will rise monotonously, which is a pure consequence of network reciprocity. It is an interesting effect, however, that defectors with high merit are less viable than low-merit defectors. It is because that the latter players can partly enjoy the neighboring cooperator-organized games but these defectors are unable to collect so high payoff which allow them to establish their own games. This phenomenon indicates implicitly how success-driven group forming could be a powerful tool because it allows defectors to be successful just for a short period.

Interestingly, the application of higher threshold values for group-formation protocol can produce a more exotic, non-monotonous $r$-dependence. This is illustrated in the bottom panel of Fig.~\ref{r}. Similarly to the previously discussed case, too small synergy factor is unable to maintain cooperation and the evolution will terminate onto a full-defector state. Nevertheless, the positive effect of success-based selection can be detected already at much smaller $r$ values. In particular, the system can evolve into a cooperator dominated state at noticeably low $r=2.59$ synergy factor, which is much lower than $r_c=3.74$ where the coexistence of $C$ and $D$ strategies starts in the traditional model. This improvement is a beautiful manifestation of the efficient selection mechanism of success-driven group formation. More precisely, the combination of a low synergy factor $r$ and a relatively high threshold value provide a special demand which can only be fulfilled by cooperator players who are surrounded by similarly cooperator neighbors. Lonely defectors may gain the requested high payoff temporarily, but the exploitation of their neighbors will weaken the cheated cooperator players who are unable to organize their own games anymore. As a consequence, the mentioned defectors cannot collect payoff from external groups anymore, hence their fitness will not be competitive to players in a $C$ domain.

The above described argument suggests that if we increase $r$ then we may manipulate the system in an undesired way because higher synergy factor makes possible for different types of players to reach the desirably high payoff. The bottom panel of Fig.~\ref{r} confirms this conclusion. Namely, by increasing $r$ the cooperation level decreases drastically because defectors can also collect the requested payoff in the vicinity of cooperators. As a straightforward consequence, increasing $r$ would further weaken the highly selected position of $C_h$ players hence we would expect lower cooperation level. But here, as we approach the critical $r_c=3.74$ value, another mechanism starts working because clustering cooperators can support each other via network reciprocity which will elevate the cooperation level in a similar fashion as we observed in the top panel of Fig.~\ref{r}. 

To gain a deeper insight about the microscopic mechanism that is responsible for the cooperator supporting group formation we present series of snapshots of the evolution at parameter values where this selection is functioning clearly. In Figure~\ref{snapshot} the simulation was started from a completely random initial state (not shown here) where all available states of players are present with equal weights. Because of the strict demands almost every players become in low-merit state after a short period, as it is illustrated in panel~(a). Here only those cooperators can fulfill the high threshold of payoff who are surrounded by akin cooperator players (they are marked by dark blue) in the snapshots. This is a crucial difference between $C$ and $D$ strategies because defectors' high payoff can only be reached on the expense of neighbors who become weak (exploited) and are unable to fulfill the demand of high threshold. Consequently, they are unable to organize their own game which also weakens indirectly the position of neighboring defectors. As panel~(a) illustrates, expect of the small cooperator islands all the other players are trapped into a low-merit state independently of they are cooperators or defectors. In the absence of high-merit players a neutral drift starts evolving between the competing $C$ and $D$ strategies because none of them can collect actual payoff. Interestingly,  the resulting coarsening of low-merit $C$ and $D$ domains also supports the growth of high-merit $C_h$ domains which can expand gradually in the growing matrix of $C_l$ players. 

It is crucial to note that the propagation of $C_h$ state in a $C_l$ domain is possible because of the stochastic character of merit's update protocol defined by Eq.~\ref{merit}. More precisely, the applied noise paves the path for a biased interface motion between $C_h$ and $C_l$ domains. Here, it is more likely that a $C_l$ player switches to $C_h$ state at the border of their domains than the reversed process because the vicinity of a high-merit neighbor can provide already a reasonable payoff for $C_l$ players. However, in case of deterministic player qualification process, or in other words in the absence of noise in merit selection, the interface separating $C_h$ and $C_l$ players would be frozen and the whole cooperator supporting effect would be less effective. In the latter case, on the one hand, the previously described neutral drift between $C$ and $D$ strategies would result in an all $C$ system for high $H$ values. On the other hand, the resulting state can hardly be considered as cooperating population because it dominantly contains $C_l$ players who cannot organize games and in the absence of common pools they cannot properly contribute.

Turning back to the noisy merit update protocol, it may also happen that a defector meets the border of a $C_h$ domain. The vicinity of active cooperators offers a chance for defectors to reach the high merit status as it is shown in panel~(c) of Fig.~\ref{snapshot}. Their success, however, is just a short term victory because it already involves the shade of their failure. First, neighboring cooperators, who provide the success of their defector fall into a low-merit status hence they are unable to organize their own games anymore. Second, they imitate the strategy of their more successful neighbor which lessens further the income of the focal defector who becomes completely vulnerable. As a result, defectors can frequently rise in the stationary state but they fall immediately which offers a narrow time window of viability to them in the sea of cooperators. A representative snapshot of the emerging morphology is shown in panel~(d) of Fig.~\ref{snapshot}.

\begin{figure}
\centerline{\epsfig{file=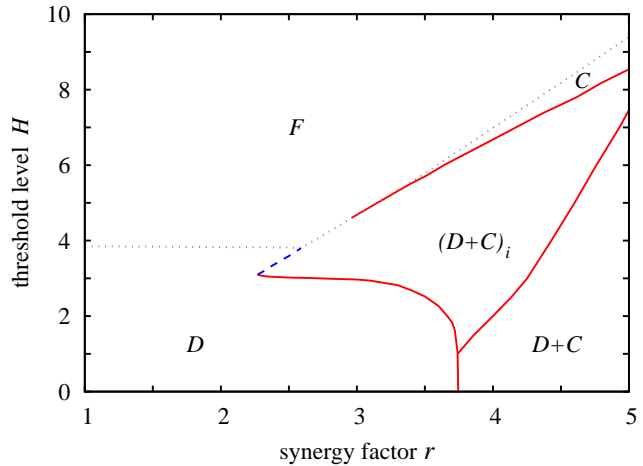,width=9.0cm}}
\caption{(Color online) Full $r-H$ phase diagram for public goods game when success-driven group formation is applied.
Solid red line denotes continuous phase transition, while dashed blue line marks discontinuous phase transition. $D$ ($C$) denotes full defector (cooperator) phase while $D+C$ marks the phase where competing strategies coexist. $(D+C)_i$ denotes a phase where the application of success-driven group organization results in a higher cooperation level comparing to the traditional model. Grey dotted line denotes the border of fixed $F$ state where high-ranking players extinct at an early stage of evolution.}
\label{phd}
\end{figure}

From the above described microscopic mechanism we can conclude that a delicate combination of $r$ and $H$ parameters is necessary to observe the positive impact of success-driven group formation protocol. To explore the complete behavior of our model we present the full $r - H$ phase diagram in Fig.~\ref{phd}. This diagram confirms our expectation, namely, if $H$ is low then the behavior of present model practically agrees with the behavior of the traditional spatial model. Namely, the system always evolves to a full $D$ state for low $r$ while the coexistence of competing strategies can be observed above a critical $r_c=3.74$ value of synergy factor, which is a straightforward consequence of network reciprocity. The latter phase is denoted by $D+C$ in the diagram. It is also in agreement with our expectation that too large $H$ value transfers the model into an uninteresting situation where all high-merit players extinct at a very early stage of the evolution and the remaining low-merit strategies ``compete" without proper interaction (they both collect zero payoff). At intermediate $H$ and $r$ values, however, we can witness how effectively the group leader selection mechanism supports cooperator strategy. We have denoted by ``improved" $(D+C)_i$ the phase where cooperators and defectors coexist but the average level of cooperation exceeds the corresponding level of traditional model when using the same $r$ value. This kind of support could be so powerful that defectors extinct and we can reach a fully cooperator phase, marked by $C$ in the diagram.

\begin{figure}
\centerline{\epsfig{file=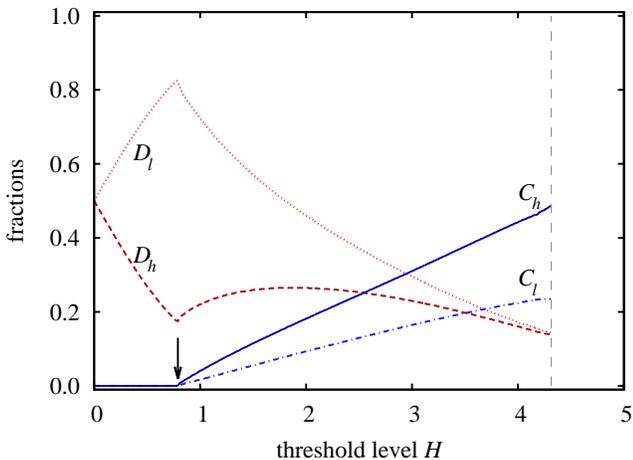,width=9.0cm}}
\caption{(Color online) Stationary fractions of the four possible states of players in dependence on threshold value $H$ as obtained for $r=3.5$ on random graph where $\langle k \rangle = 4$ uniform degree distribution was applied.  
If threshold $H$ is increased above a critical value, marked by an arrow, then success-driven group formation mechanism is capable to elevate the cooperation level significantly. Above a critical value of $H$, denoted by a dashed line, the demand towards group organizers becomes so high that nobody can fulfill it and the system is trapped into a  state of low-merit players.}
\label{random}
\end{figure}

We would like to emphasize that the positive impact of success-driven group formation protocol is not restricted to the applied topology, but remains valid if we use less regular interaction graphs. To illustrate it we present results for stationary states obtained on a random graph where uniform degree distribution is preserved and every node has  $\langle k \rangle =4$ neighbors \cite{szabo_jpa04}. This modification allows us to introduce only randomness of links without involving additional disturbing effects. As Figure~\ref{random} shows the cooperation level starts raising if threshold exceeds $H_c=0.79$ value which is marked by an arrow in the plot. By increasing $H$ further the cooperator supporting selection mechanism becomes more effective and the sum of cooperator player grows monotonously. This positive impact terminates only if $H$ becomes too high because $H>4.32$ is proved to be inaccessible demand for groups organizers at this value of $r$. Here, as we have already discussed for square lattice topology, the initial high-merit players die out immediately after the evolution is launched and only low-merit players remain. 

\begin{figure}
\centerline{\epsfig{file=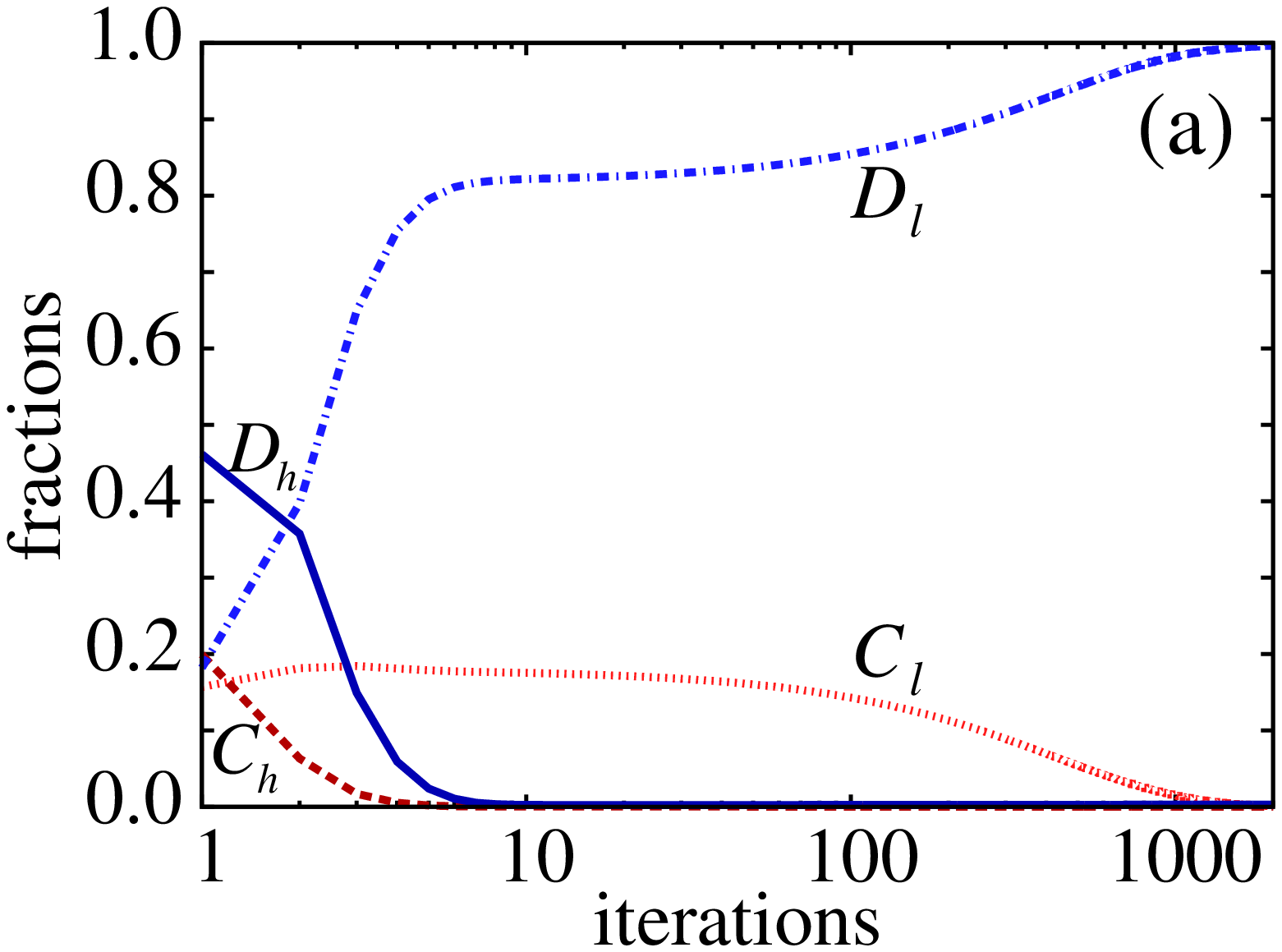,width=4.5cm}\epsfig{file=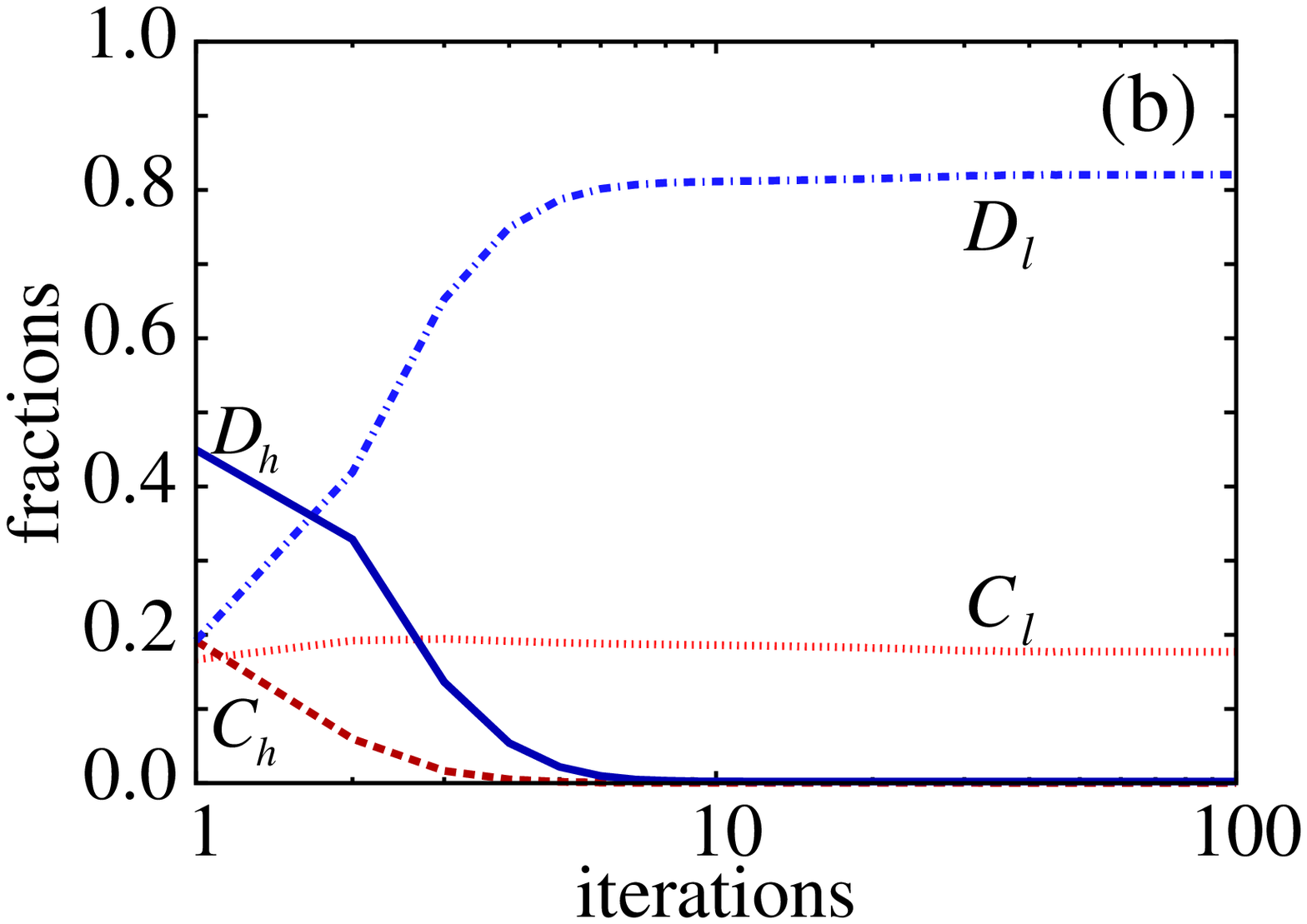,width=4.4cm}}
\centerline{\epsfig{file=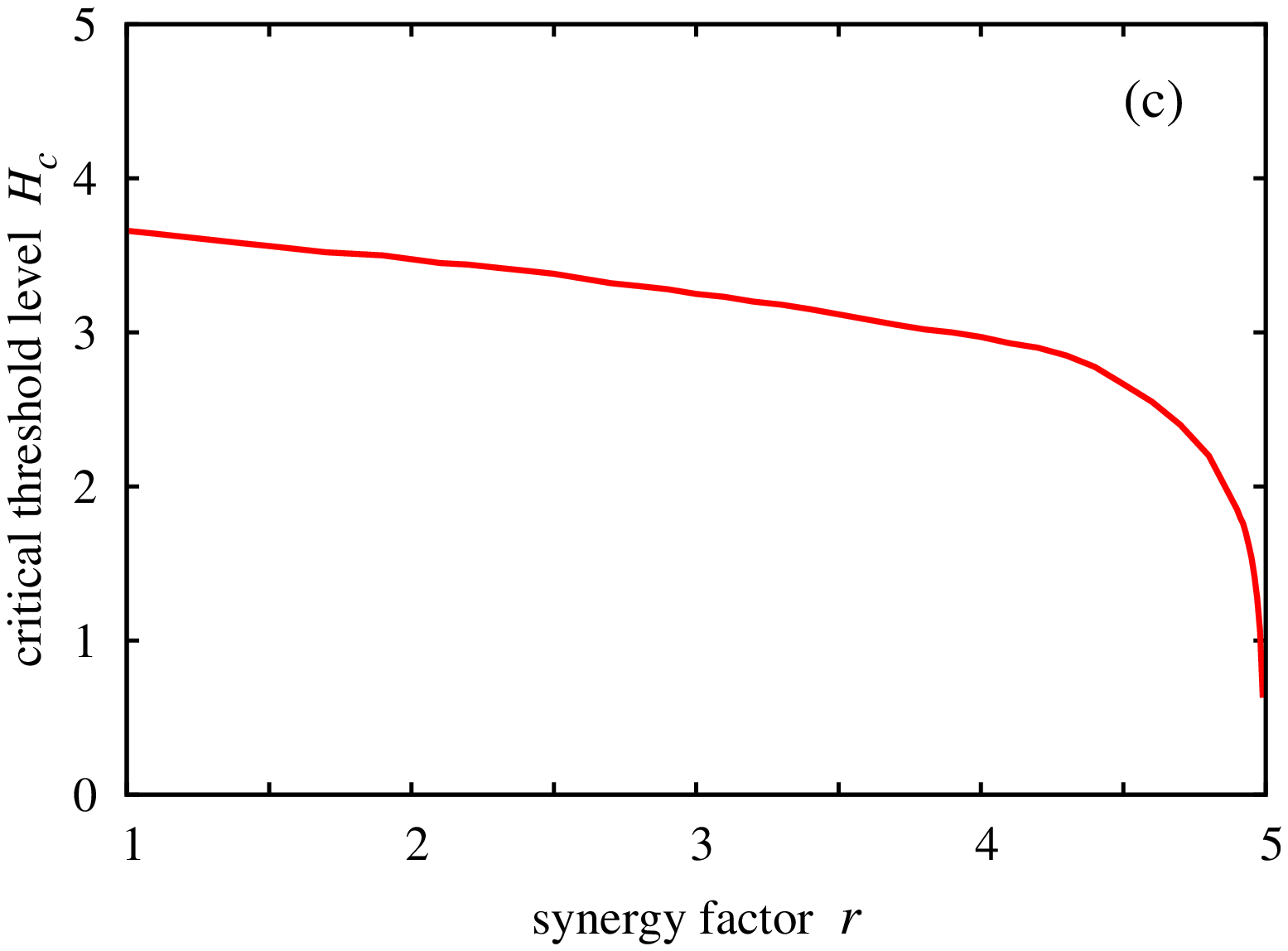,width=8.0cm}}
\caption{(Color online) Success-driven group formation in well-mixed populations. Top panels show how the fractions of players' state change when the system evolves from a random initial state for different threshold values of $H$. If $H$ is low, shown in panel~(a), then all cooperators die out and the system terminates into a full-defector state. If $H$ is high enough, shown in panel~(b), then all high-merit players extinct independently of their strategies and only low-merit players remain who are unable to organize groups and play public goods game. Threshold values are $H=2.95$ and 3.0 respectively, while $r=4.0$ for both cases. Panel~(c) depicts the critical value of $H$ in dependence on synergy factor which is the border line of the above mentioned two possible trajectories.}
\label{mf}
\end{figure}

The results summarized in Fig.~\ref{random} supports our argument that the interaction topology has second order importance because the proposed selection mechanism will always amplify the positive consequence of network reciprocity. The only crucial criterion is to have stable partners during the strategy evolution. We can easily confirm this argument indirectly if we consider the well-mixed version of the proposed model. In the latter case network reciprocity cannot function because players have no limited and stable connections. 

To check the differences we can also consider the case of unstructured population and solve the related replicator equations numerically. Details of this calculation are given in the Appendix while the main results are summarized in Fig.~\ref{mf}. As the top two panels illustrate, there are two significantly different trajectories if we launch the evolution from a random initial state where all states are present with equal weights. If the applied threshold level is low, shown in panel~(a), then all cooperators will die out no matter if they have high or low initial merit. If $H$ is above a critical $H_c$ value then another two states extinct. Namely, both $C_h$ and $D_h$ players die out and the system is trapped in a state where only low-merit players are present hence there is no proper interaction between competing strategies. Naturally, the actual value of $H_c$ where these trajectories change depends on the value of synergy factor $r$, as it is summarized in panel~(c). The latter plot suggests that in case of harsh conditions, when $r$ is low, a higher threshold value is necessary to avoid the full-$D$ state because low $r$ values offer an obvious advantage for defectors, which will provide a fast extinction of cooperators. This time course can be avoided only if $H$ is so high that it provides an unsolvable barrier to defectors as well, hence they cannot organize their groups and cannot collect payoff anymore. Nevertheless, the presented results support our expectation, namely, the application of success-driven group formation protocol can support cooperation actively only in structured populations where it provides an elegant way to amplify network reciprocity.

\section{Discussion}

It is a widespread observation that people are not always enthusiastic to participate in a joint venture because the one who establishes it could have a decisive role in its success. In particular, if the one who organizes the game is unsuccessful in general then neighbors are reluctant to join, while a successful actor is always an attractive target of investments. Motivated by these observations we proposed a minimal model where only those players can establish a group and announce a public goods game whose previous payoff exceeds a threshold level. Otherwise, a player who fails to fulfill this criterion can only participate in other's game. 

Our results demonstrated that the proposed protocol can support cooperation in structured populations effectively, because individual success can only be maintained if a player tries to cooperate with its neighbors, while a defector's success could only be short-lived because it is based on the exploitation of others. In this way success, which is principally a strategy-neutral demand and cannot be told its origin in advance, could be a powerful selection mechanism which reconciles individual and collective interests. 

In other words, success serves as an individual reputation that can inform neighbors how to interact with other potential partners. In contrast to the traditional assumption about reputation, however, we do not suppose a priori a positive behavior about a player when high-reputation is considered \cite{milinski_n02,brandt_prsb03,fu_pre08b}. Still, success can only work permanently for those players who are responsible to their neighborhood's success as well. This observation fits nicely to more general findings where individual success was a key factor of strategy update \cite{helbing_pnas09, perc_pre11, yang_zh_epjb13, chen_xj_sr16}.

Interestingly, higher demand for success is proved to be more effective selection criterion for a higher cooperation level. This observation conceptually fits to some previous findings when harsh environment, which is modeled by less attractive payoff parameters of the actual social game, resulted in higher cooperation level in the population \cite{smaldino_csf13,szolnoki_epl14}. In our present case the explanation of this seemingly counterintuitive behavior is that defectors can only reach a limited level of success because really high payoff can only be gained with the help of successful associates.

This research was supported by the Hungarian National Research Fund (Grant K-120785), the National Natural Science Foundation of China (Grants No. 61503062), and the Fundamental Research Funds of the Central Universities of China.

\section*{Appendix}

In the following we summarize the replicator equations \cite{hofbauer_98} and related payoff values for well-mixed population when success-driven group formation protocol is applied. Here the fraction of $D_l, D_h, C_l$, and $C_h$ players are denoted by $x_{_{D_l}}, x_{_{D_h}}, x_{_{C_l}}$, and $x_{_{C_h}}$ respectively. Because of their total numbers are fixed their fractions fulfill the equation $x_{_{D_l}}+x_{_{D_h}}+x_{_{C_l}}+x_{_{C_h}}=1$. The related equations which describe the rates of changes of the strategies are as follows:
\begin{eqnarray}
\dot{x}_{_{C_h}} &=& x_{_{C_h}} (\Pi_{_{C_h}} - \overline{\Pi}_h) + x_{_{C_l}} m(\Pi_{_{C_l}}) - x_{_{C_h}} [1-m(\Pi_{_{C_h}})] \nonumber\\
\dot{x}_{_{C_l}} &=& x_{_{C_l}} (\Pi_{_{C_l}} - \overline{\Pi}_l) + x_{_{C_h}} [1- m(\Pi_{_{C_h}})] - x_{_{C_l}} m(\Pi_{_{C_l}}) \nonumber\\
\dot{x}_{_{D_l}} &=& x_{_{D_l}} (\Pi_{_{D_l}} - \overline{\Pi}_l) +
x_{_{D_h}} [1- m(\Pi_{_{D_h}})] - x_{_{D_l}} m(\Pi_{_{D_l}}).
\nonumber \label{replicator}
\end{eqnarray}
Here dots denote the derivatives with respect to time $t$ and $m(z)$ is the well-known Fermi-function which dictates how individual merit evolves according to Eq.~\ref{merit}. The first term in the equations characterizes the rate of changes of strategies by strategy updating, and the second and third terms in the equations characterize the rate of changes of strategies by merit updating. The averaged payoff $\overline{\Pi}_h$ in the subpopulation with the high merit value and the averaged payoff $\overline{\Pi}_l$ in the subpopulation with the low merit value are respectively given by
\begin{equation}
\overline{\Pi}_h = (x_{_{C_h}} \Pi_{_{C_h}} + x_{_{D_h}}
\Pi_{_{D_h}})/(x_{_{C_h}}+x_{_{D_h}}) \,, \label{averP}
\end{equation}
and
\begin{equation}
\overline{\Pi}_l = (x_{_{C_l}} \Pi_{_{C_l}} + x_{_{D_l}}
\Pi_{_{D_l}})/(x_{_{C_l}}+x_{_{D_l}}) \,, \label{averP}
\end{equation}
where the average payoff for each subset of players are:
\begin{widetext}
\begin{eqnarray}
\Pi_{_{C_h}} =& & \sum_{i+j+k+n=N-1} \frac{(N-1)!}{i!j!k!n!}x_{_{C_l}}^i x_{_{C_h}}^j x_{_{D_l}}^k x_{_{D_h}}^n \left( \frac{r(i+j+1)}{N} - 1 \right) +  \nonumber\\
& & \sum_{i+j+k+n=N-2} \frac{(N-2)!}{i!j!k!n!}(N-1)x_{_{C_h}}x_{_{C_l}}^i x_{_{C_h}}^j x_{_{D_l}}^k x_{_{D_h}}^n \left( \frac{r(i+j+2)}{N} - 1 \right) + \nonumber \\
& & \sum_{i+j+k+n=N-2} \frac{(N-2)!}{i!j!k!n!}(N-1)x_{_{D_h}}x_{_{C_l}}^i x_{_{C_h}}^j x_{_{D_l}}^k x_{_{D_h}}^n \left( \frac{r(i+j+1)}{N} - 1 \right)   \nonumber\\
\Pi_{_{C_l}} =& & \sum_{i+j+k+n=N-2} \frac{(N-2)!}{i!j!k!n!}(N-1)x_{_{C_h}}x_{_{C_l}}^i x_{_{C_h}}^j x_{_{D_l}}^k x_{_{D_h}}^n \left( \frac{r(i+j+2)}{N} - 1 \right) + \nonumber \\
& & \sum_{i+j+k+n=N-2} \frac{(N-2)!}{i!j!k!n!}(N-1)x_{_{D_h}}x_{_{C_l}}^i x_{_{C_h}}^j x_{_{D_l}}^k x_{_{D_h}}^n \left( \frac{r(i+j+1)}{N} - 1 \right)   \nonumber\\
\Pi_{_{D_h}} =& & \sum_{i+j+k+n=N-1} \frac{(N-1)!}{i!j!k!n!}x_{_{C_l}}^i x_{_{C_h}}^j x_{_{D_l}}^k x_{_{D_h}}^n \frac{r(i+j)}{N} +  \nonumber\\
& & \sum_{i+j+k+n=N-2} \frac{(N-2)!}{i!j!k!n!}(N-1)x_{_{C_h}}x_{_{C_l}}^i x_{_{C_h}}^j x_{_{D_l}}^k x_{_{D_h}}^n \frac{r(i+j+1)}{N} + \nonumber \\
& & \sum_{i+j+k+n=N-2} \frac{(N-2)!}{i!j!k!n!}(N-1)x_{_{D_h}}x_{_{C_l}}^i x_{_{C_h}}^j x_{_{D_l}}^k x_{_{D_h}}^n \frac{r(i+j)}{N} \nonumber\\
\Pi_{_{D_l}} =& & \sum_{i+j+k+n=N-2} \frac{(N-2)!}{i!j!k!n!}(N-1)x_{_{C_h}}x_{_{C_l}}^i x_{_{C_h}}^j x_{_{D_l}}^k x_{_{D_h}}^n \frac{r(i+j+1)}{N} + \nonumber \\
& & \sum_{i+j+k+n=N-2} \frac{(N-2)!}{i!j!k!n!}(N-1)x_{_{D_h}}x_{_{C_l}}^i x_{_{C_h}}^j x_{_{D_l}}^k x_{_{D_h}}^n \frac{r(i+j)}{N}  \nonumber\\
\end{eqnarray}
\end{widetext}


\begin{thebibliography}{51}
\expandafter\ifx\csname natexlab\endcsname\relax\def\natexlab#1{#1}\fi
\expandafter\ifx\csname bibnamefont\endcsname\relax
  \def\bibnamefont#1{#1}\fi
\expandafter\ifx\csname bibfnamefont\endcsname\relax
  \def\bibfnamefont#1{#1}\fi
\expandafter\ifx\csname citenamefont\endcsname\relax
  \def\citenamefont#1{#1}\fi
\expandafter\ifx\csname url\endcsname\relax
  \def\url#1{\texttt{#1}}\fi
\expandafter\ifx\csname urlprefix\endcsname\relax\def\urlprefix{URL }\fi
\providecommand{\bibinfo}[2]{#2}
\providecommand{\eprint}[2][]{\url{#2}}

\bibitem[{\citenamefont{Sigmund}(2010)}]{sigmund_10}
\bibinfo{author}{\bibfnamefont{K.}~\bibnamefont{Sigmund}},
  \emph{\bibinfo{title}{The Calculus of Selfishness}}
  (\bibinfo{publisher}{Princeton University Press},
  \bibinfo{address}{Princeton, NJ}, \bibinfo{year}{2010}).

\bibitem[{\citenamefont{Hardin}(1968)}]{hardin_g_s68}
\bibinfo{author}{\bibfnamefont{G.}~\bibnamefont{Hardin}},
  \bibinfo{journal}{Science} \textbf{\bibinfo{volume}{162}},
  \bibinfo{pages}{1243} (\bibinfo{year}{1968}).

\bibitem[{\citenamefont{Gibbons}(1992)}]{gibbons_92}
\bibinfo{author}{\bibfnamefont{R.}~\bibnamefont{Gibbons}},
  \emph{\bibinfo{title}{Game Theory for Applied Economists}}
  (\bibinfo{publisher}{Princeton University Press},
  \bibinfo{address}{Princeton}, \bibinfo{year}{1992}).

\bibitem[{\citenamefont{Ledyard}(1997)}]{ledyard_97}
\bibinfo{author}{\bibfnamefont{J.~O.} \bibnamefont{Ledyard}}, in
  \emph{\bibinfo{booktitle}{The Handbook of Experimental Economics}}, edited by
  \bibinfo{editor}{\bibfnamefont{J.~H.} \bibnamefont{Kagel}} \bibnamefont{and}
  \bibinfo{editor}{\bibfnamefont{A.~E.} \bibnamefont{Roth}}
  (\bibinfo{publisher}{Princeton University Press},
  \bibinfo{address}{Princeton, NJ}, \bibinfo{year}{1997}), pp.
  \bibinfo{pages}{111--194}.

\bibitem[{\citenamefont{Nowak and Highfield}(2011)}]{nowak_11}
\bibinfo{author}{\bibfnamefont{M.~A.} \bibnamefont{Nowak}} \bibnamefont{and}
  \bibinfo{author}{\bibfnamefont{R.}~\bibnamefont{Highfield}},
  \emph{\bibinfo{title}{SuperCooperators: Altruism, Evolution, and Why We Need
  Each Other to Succeed}} (\bibinfo{publisher}{Free Press},
  \bibinfo{address}{New York}, \bibinfo{year}{2011}).

\bibitem[{\citenamefont{Sicardi et~al.}(2009)\citenamefont{Sicardi, Fort,
  Vainstein, and Arenzon}}]{sicardi_jtb09}
\bibinfo{author}{\bibfnamefont{E.~A.} \bibnamefont{Sicardi}},
  \bibinfo{author}{\bibfnamefont{H.}~\bibnamefont{Fort}},
  \bibinfo{author}{\bibfnamefont{M.~H.} \bibnamefont{Vainstein}},
  \bibnamefont{and} \bibinfo{author}{\bibfnamefont{J.~J.}
  \bibnamefont{Arenzon}}, \bibinfo{journal}{J. Theor. Biol.}
  \textbf{\bibinfo{volume}{256}}, \bibinfo{pages}{240} (\bibinfo{year}{2009}).

\bibitem[{\citenamefont{Fu et~al.}(2012)\citenamefont{Fu, Tarnita, Christakis,
  Wang, Rand, and Nowak}}]{fu_srep12}
\bibinfo{author}{\bibfnamefont{F.}~\bibnamefont{Fu}},
  \bibinfo{author}{\bibfnamefont{C.}~\bibnamefont{Tarnita}},
  \bibinfo{author}{\bibfnamefont{N.}~\bibnamefont{Christakis}},
  \bibinfo{author}{\bibfnamefont{L.}~\bibnamefont{Wang}},
  \bibinfo{author}{\bibfnamefont{D.}~\bibnamefont{Rand}}, \bibnamefont{and}
  \bibinfo{author}{\bibfnamefont{M.}~\bibnamefont{Nowak}},
  \bibinfo{journal}{Sci. Rep.} \textbf{\bibinfo{volume}{2}},
  \bibinfo{pages}{460} (\bibinfo{year}{2012}).

\bibitem[{\citenamefont{Rezaei and Kirley}(2012)}]{rezaei_pa12}
\bibinfo{author}{\bibfnamefont{G.}~\bibnamefont{Rezaei}} \bibnamefont{and}
  \bibinfo{author}{\bibfnamefont{M.}~\bibnamefont{Kirley}},
  \bibinfo{journal}{Physica A} \textbf{\bibinfo{volume}{391}},
  \bibinfo{pages}{6199} (\bibinfo{year}{2012}).

\bibitem[{\citenamefont{Pe{\~n}a et~al.}(2009)\citenamefont{Pe{\~n}a, Volken,
  Pestelacci, and Tomassini}}]{pena_pre09}
\bibinfo{author}{\bibfnamefont{J.}~\bibnamefont{Pe{\~n}a}},
  \bibinfo{author}{\bibfnamefont{H.}~\bibnamefont{Volken}},
  \bibinfo{author}{\bibfnamefont{E.}~\bibnamefont{Pestelacci}},
  \bibnamefont{and}
  \bibinfo{author}{\bibfnamefont{M.}~\bibnamefont{Tomassini}},
  \bibinfo{journal}{Phys. Rev. E} \textbf{\bibinfo{volume}{80}},
  \bibinfo{pages}{016110} (\bibinfo{year}{2009}).

\bibitem[{\citenamefont{Arenas et~al.}(2011)\citenamefont{Arenas, Camacho,
  Cuesta, and Requejo}}]{arenas_jtb11}
\bibinfo{author}{\bibfnamefont{A.}~\bibnamefont{Arenas}},
  \bibinfo{author}{\bibfnamefont{J.}~\bibnamefont{Camacho}},
  \bibinfo{author}{\bibfnamefont{J.~A.} \bibnamefont{Cuesta}},
  \bibnamefont{and} \bibinfo{author}{\bibfnamefont{R.}~\bibnamefont{Requejo}},
  \bibinfo{journal}{J. Theor. Biol.} \textbf{\bibinfo{volume}{279}},
  \bibinfo{pages}{113} (\bibinfo{year}{2011}).

\bibitem[{\citenamefont{Mobilia}(2012)}]{mobilia_pre12}
\bibinfo{author}{\bibfnamefont{M.}~\bibnamefont{Mobilia}},
  \bibinfo{journal}{Phys. Rev. E.} \textbf{\bibinfo{volume}{86}},
  \bibinfo{pages}{011134} (\bibinfo{year}{2012}).

\bibitem[{\citenamefont{Wu et~al.}(2009)\citenamefont{Wu, Fu, and
  Wang}}]{wu_t_pre09}
\bibinfo{author}{\bibfnamefont{T.}~\bibnamefont{Wu}},
  \bibinfo{author}{\bibfnamefont{F.}~\bibnamefont{Fu}}, \bibnamefont{and}
  \bibinfo{author}{\bibfnamefont{L.}~\bibnamefont{Wang}},
  \bibinfo{journal}{Phys. Rev. E} \textbf{\bibinfo{volume}{80}},
  \bibinfo{pages}{026121} (\bibinfo{year}{2009}).

\bibitem[{\citenamefont{Requejo et~al.}(2012)\citenamefont{Requejo, Camacho,
  Cuesta, and Arenas}}]{requejo_pre12b}
\bibinfo{author}{\bibfnamefont{R.~J.} \bibnamefont{Requejo}},
  \bibinfo{author}{\bibfnamefont{J.}~\bibnamefont{Camacho}},
  \bibinfo{author}{\bibfnamefont{J.~A.} \bibnamefont{Cuesta}},
  \bibnamefont{and} \bibinfo{author}{\bibfnamefont{A.}~\bibnamefont{Arenas}},
  \bibinfo{journal}{Phys. Rev. E} \textbf{\bibinfo{volume}{86}},
  \bibinfo{pages}{026105} (\bibinfo{year}{2012}).

\bibitem[{\citenamefont{Sasaki et~al.}(2007)\citenamefont{Sasaki, Okada, and
  Unemi}}]{sasaki_prsb07}
\bibinfo{author}{\bibfnamefont{T.}~\bibnamefont{Sasaki}},
  \bibinfo{author}{\bibfnamefont{I.}~\bibnamefont{Okada}}, \bibnamefont{and}
  \bibinfo{author}{\bibfnamefont{T.}~\bibnamefont{Unemi}},
  \bibinfo{journal}{Proc. R. Soc. Lond. B} \textbf{\bibinfo{volume}{274}},
  \bibinfo{pages}{2639} (\bibinfo{year}{2007}).

\bibitem[{\citenamefont{Liu et~al.}(2013)\citenamefont{Liu, Chen, Zhang, Tao,
  and Wang}}]{liu_yk_epl13}
\bibinfo{author}{\bibfnamefont{Y.}~\bibnamefont{Liu}},
  \bibinfo{author}{\bibfnamefont{X.}~\bibnamefont{Chen}},
  \bibinfo{author}{\bibfnamefont{L.}~\bibnamefont{Zhang}},
  \bibinfo{author}{\bibfnamefont{F.}~\bibnamefont{Tao}}, \bibnamefont{and}
  \bibinfo{author}{\bibfnamefont{L.}~\bibnamefont{Wang}},
  \bibinfo{journal}{EPL} \textbf{\bibinfo{volume}{102}}, \bibinfo{pages}{50006}
  (\bibinfo{year}{2013}).

\bibitem[{\citenamefont{Vilone et~al.}(2011)\citenamefont{Vilone, S{\'a}nchez,
  and G{\'o}mez-Garde{\~n}es}}]{vilone_jsm11}
\bibinfo{author}{\bibfnamefont{D.}~\bibnamefont{Vilone}},
  \bibinfo{author}{\bibfnamefont{A.}~\bibnamefont{S{\'a}nchez}},
  \bibnamefont{and}
  \bibinfo{author}{\bibfnamefont{J.}~\bibnamefont{G{\'o}mez-Garde{\~n}es}},
  \bibinfo{journal}{J. Stat. Mech} \textbf{\bibinfo{volume}{2011}},
  \bibinfo{pages}{P04019} (\bibinfo{year}{2011}).

\bibitem[{\citenamefont{Hindersin and Traulsen}(2015)}]{hindersin_pcbi15}
\bibinfo{author}{\bibfnamefont{L.}~\bibnamefont{Hindersin}} \bibnamefont{and}
  \bibinfo{author}{\bibfnamefont{A.}~\bibnamefont{Traulsen}},
  \bibinfo{journal}{PLoS Comput. Biol.} \textbf{\bibinfo{volume}{11}},
  \bibinfo{pages}{e1004437} (\bibinfo{year}{2015}).

\bibitem[{\citenamefont{Suzuki et~al.}(2008)\citenamefont{Suzuki, Kato, and
  Arita}}]{suzuki_r_pre08}
\bibinfo{author}{\bibfnamefont{R.}~\bibnamefont{Suzuki}},
  \bibinfo{author}{\bibfnamefont{M.}~\bibnamefont{Kato}}, \bibnamefont{and}
  \bibinfo{author}{\bibfnamefont{T.}~\bibnamefont{Arita}},
  \bibinfo{journal}{Phys. Rev. E} \textbf{\bibinfo{volume}{77}},
  \bibinfo{pages}{021911} (\bibinfo{year}{2008}).

\bibitem[{\citenamefont{Mobilia}(2013)}]{mobilia_csf13}
\bibinfo{author}{\bibfnamefont{M.}~\bibnamefont{Mobilia}},
  \bibinfo{journal}{Chaos, Solitons \& Fractals} \textbf{\bibinfo{volume}{56}},
  \bibinfo{pages}{113} (\bibinfo{year}{2013}).

\bibitem[{\citenamefont{Javarone}(2016)}]{javarone_epjb16}
\bibinfo{author}{\bibfnamefont{M.~A.} \bibnamefont{Javarone}},
  \bibinfo{journal}{Eur. Phys. J. B} \textbf{\bibinfo{volume}{89}},
  \bibinfo{pages}{42} (\bibinfo{year}{2016}).

\bibitem[{\citenamefont{Jordan et~al.}(2016)\citenamefont{Jordan, Hoffman,
  Nowak, and Rand}}]{jordan_pnas16}
\bibinfo{author}{\bibfnamefont{J.~J.} \bibnamefont{Jordan}},
  \bibinfo{author}{\bibfnamefont{M.}~\bibnamefont{Hoffman}},
  \bibinfo{author}{\bibfnamefont{M.~A.} \bibnamefont{Nowak}}, \bibnamefont{and}
  \bibinfo{author}{\bibfnamefont{D.~G.} \bibnamefont{Rand}},
  \bibinfo{journal}{Proc. Natl. Acad. Sci. USA} \textbf{\bibinfo{volume}{113}},
  \bibinfo{pages}{8658} (\bibinfo{year}{2016}).

\bibitem[{\citenamefont{Press and Dyson}(2012)}]{press_pnas12}
\bibinfo{author}{\bibfnamefont{W.}~\bibnamefont{Press}} \bibnamefont{and}
  \bibinfo{author}{\bibfnamefont{F.}~\bibnamefont{Dyson}},
  \bibinfo{journal}{Proc. Natl. Acad. Sci. USA} \textbf{\bibinfo{volume}{109}},
  \bibinfo{pages}{10409} (\bibinfo{year}{2012}).

\bibitem[{\citenamefont{Burton-Chellew
  et~al.}(2016)\citenamefont{Burton-Chellew, Mouden, and
  West}}]{burton-chellew_pnas16}
\bibinfo{author}{\bibfnamefont{M.~N.} \bibnamefont{Burton-Chellew}},
  \bibinfo{author}{\bibfnamefont{C.~E.} \bibnamefont{Mouden}},
  \bibnamefont{and} \bibinfo{author}{\bibfnamefont{S.~A.} \bibnamefont{West}},
  \bibinfo{journal}{Proc. Natl. Acad. Sci. USA} \textbf{\bibinfo{volume}{113}},
  \bibinfo{pages}{1291} (\bibinfo{year}{2016}).

\bibitem[{\citenamefont{Szolnoki and Perc}(2012)}]{szolnoki_pre12}
\bibinfo{author}{\bibfnamefont{A.}~\bibnamefont{Szolnoki}} \bibnamefont{and}
  \bibinfo{author}{\bibfnamefont{M.}~\bibnamefont{Perc}},
  \bibinfo{journal}{Phys. Rev. E} \textbf{\bibinfo{volume}{85}},
  \bibinfo{pages}{026104} (\bibinfo{year}{2012}).

\bibitem[{\citenamefont{Chen et~al.}(2012)\citenamefont{Chen, Schick, Doebeli,
  Blachford, and Wang}}]{chen_xj_pone12b}
\bibinfo{author}{\bibfnamefont{X.}~\bibnamefont{Chen}},
  \bibinfo{author}{\bibfnamefont{A.}~\bibnamefont{Schick}},
  \bibinfo{author}{\bibfnamefont{M.}~\bibnamefont{Doebeli}},
  \bibinfo{author}{\bibfnamefont{A.}~\bibnamefont{Blachford}},
  \bibnamefont{and} \bibinfo{author}{\bibfnamefont{L.}~\bibnamefont{Wang}},
  \bibinfo{journal}{PLoS ONE} \textbf{\bibinfo{volume}{7}},
  \bibinfo{pages}{e36260} (\bibinfo{year}{2012}).

\bibitem[{\citenamefont{Szolnoki and Chen}(2015)}]{szolnoki_pre15}
\bibinfo{author}{\bibfnamefont{A.}~\bibnamefont{Szolnoki}} \bibnamefont{and}
  \bibinfo{author}{\bibfnamefont{X.}~\bibnamefont{Chen}},
  \bibinfo{journal}{Phys. Rev. E} \textbf{\bibinfo{volume}{92}},
  \bibinfo{pages}{042813} (\bibinfo{year}{2015}).

\bibitem[{\citenamefont{Szolnoki and Perc}(2016)}]{szolnoki_njp16}
\bibinfo{author}{\bibfnamefont{A.}~\bibnamefont{Szolnoki}} \bibnamefont{and}
  \bibinfo{author}{\bibfnamefont{M.}~\bibnamefont{Perc}},
  \bibinfo{journal}{New J. Phys.} \textbf{\bibinfo{volume}{18}},
  \bibinfo{pages}{083021} (\bibinfo{year}{2016}).

\bibitem[{\citenamefont{Hauert et~al.}(2002)\citenamefont{Hauert, De~Monte,
  Hofbauer, and Sigmund}}]{hauert_s02}
\bibinfo{author}{\bibfnamefont{C.}~\bibnamefont{Hauert}},
  \bibinfo{author}{\bibfnamefont{S.}~\bibnamefont{De~Monte}},
  \bibinfo{author}{\bibfnamefont{J.}~\bibnamefont{Hofbauer}}, \bibnamefont{and}
  \bibinfo{author}{\bibfnamefont{K.}~\bibnamefont{Sigmund}},
  \bibinfo{journal}{Science} \textbf{\bibinfo{volume}{296}},
  \bibinfo{pages}{1129} (\bibinfo{year}{2002}).

\bibitem[{\citenamefont{Semmann et~al.}(2003)\citenamefont{Semmann, Krambeck,
  and Milinski}}]{semmann_n03}
\bibinfo{author}{\bibfnamefont{D.}~\bibnamefont{Semmann}},
  \bibinfo{author}{\bibfnamefont{H.-J.} \bibnamefont{Krambeck}},
  \bibnamefont{and} \bibinfo{author}{\bibfnamefont{M.}~\bibnamefont{Milinski}},
  \bibinfo{journal}{Nature} \textbf{\bibinfo{volume}{425}},
  \bibinfo{pages}{390} (\bibinfo{year}{2003}).

\bibitem[{\citenamefont{Nowak and May}(1992)}]{nowak_n92b}
\bibinfo{author}{\bibfnamefont{M.~A.} \bibnamefont{Nowak}} \bibnamefont{and}
  \bibinfo{author}{\bibfnamefont{R.~M.} \bibnamefont{May}},
  \bibinfo{journal}{Nature} \textbf{\bibinfo{volume}{359}},
  \bibinfo{pages}{826} (\bibinfo{year}{1992}).

\bibitem[{\citenamefont{Santos and Pacheco}(2005)}]{santos_prl05}
\bibinfo{author}{\bibfnamefont{F.~C.} \bibnamefont{Santos}} \bibnamefont{and}
  \bibinfo{author}{\bibfnamefont{J.~M.} \bibnamefont{Pacheco}},
  \bibinfo{journal}{Phys. Rev. Lett.} \textbf{\bibinfo{volume}{95}},
  \bibinfo{pages}{098104} (\bibinfo{year}{2005}).

\bibitem[{\citenamefont{Szab{\'o} and F{\'a}th}(2007)}]{szabo_pr07}
\bibinfo{author}{\bibfnamefont{G.}~\bibnamefont{Szab{\'o}}} \bibnamefont{and}
  \bibinfo{author}{\bibfnamefont{G.}~\bibnamefont{F{\'a}th}},
  \bibinfo{journal}{Phys. Rep.} \textbf{\bibinfo{volume}{446}},
  \bibinfo{pages}{97} (\bibinfo{year}{2007}).

\bibitem[{\citenamefont{Nowak}(2006)}]{nowak_s06}
\bibinfo{author}{\bibfnamefont{M.~A.} \bibnamefont{Nowak}},
  \bibinfo{journal}{Science} \textbf{\bibinfo{volume}{314}},
  \bibinfo{pages}{1560} (\bibinfo{year}{2006}).

\bibitem[{\citenamefont{Szolnoki and Szab{\'o}}(2007)}]{szolnoki_epl07}
\bibinfo{author}{\bibfnamefont{A.}~\bibnamefont{Szolnoki}} \bibnamefont{and}
  \bibinfo{author}{\bibfnamefont{G.}~\bibnamefont{Szab{\'o}}},
  \bibinfo{journal}{EPL} \textbf{\bibinfo{volume}{77}}, \bibinfo{pages}{30004}
  (\bibinfo{year}{2007}).

\bibitem[{\citenamefont{Perc et~al.}(2008)\citenamefont{Perc, Szolnoki, and
  Szab{\'o}}}]{perc_pre08b}
\bibinfo{author}{\bibfnamefont{M.}~\bibnamefont{Perc}},
  \bibinfo{author}{\bibfnamefont{A.}~\bibnamefont{Szolnoki}}, \bibnamefont{and}
  \bibinfo{author}{\bibfnamefont{G.}~\bibnamefont{Szab{\'o}}},
  \bibinfo{journal}{Phys. Rev. E} \textbf{\bibinfo{volume}{78}},
  \bibinfo{pages}{066101} (\bibinfo{year}{2008}).

\bibitem[{\citenamefont{Wang et~al.}(2012)\citenamefont{Wang, Szolnoki, and
  Perc}}]{wang_z_epl12}
\bibinfo{author}{\bibfnamefont{Z.}~\bibnamefont{Wang}},
  \bibinfo{author}{\bibfnamefont{A.}~\bibnamefont{Szolnoki}}, \bibnamefont{and}
  \bibinfo{author}{\bibfnamefont{M.}~\bibnamefont{Perc}},
  \bibinfo{journal}{EPL} \textbf{\bibinfo{volume}{97}}, \bibinfo{pages}{48001}
  (\bibinfo{year}{2012}).

\bibitem[{\citenamefont{G{\'o}mez-Garde{\~n}es
  et~al.}(2012)\citenamefont{G{\'o}mez-Garde{\~n}es, Reinares, Arenas, and
  Flor{\' \i}a}}]{gomez-gardenes_srep12}
\bibinfo{author}{\bibfnamefont{J.}~\bibnamefont{G{\'o}mez-Garde{\~n}es}},
  \bibinfo{author}{\bibfnamefont{I.}~\bibnamefont{Reinares}},
  \bibinfo{author}{\bibfnamefont{A.}~\bibnamefont{Arenas}}, \bibnamefont{and}
  \bibinfo{author}{\bibfnamefont{L.~M.} \bibnamefont{Flor{\' \i}a}},
  \bibinfo{journal}{Sci. Rep.} \textbf{\bibinfo{volume}{2}},
  \bibinfo{pages}{620} (\bibinfo{year}{2012}).

\bibitem[{\citenamefont{Perc et~al.}(2013)\citenamefont{Perc,
  G{\'o}mez-Garde{\~n}es, Szolnoki, and Flor{\'{\i}a and Y.
  Moreno}}}]{perc_jrsi13}
\bibinfo{author}{\bibfnamefont{M.}~\bibnamefont{Perc}},
  \bibinfo{author}{\bibfnamefont{J.}~\bibnamefont{G{\'o}mez-Garde{\~n}es}},
  \bibinfo{author}{\bibfnamefont{A.}~\bibnamefont{Szolnoki}}, \bibnamefont{and}
  \bibinfo{author}{\bibfnamefont{L.~M.} \bibnamefont{Flor{\'{\i}a and Y.
  Moreno}}}, \bibinfo{journal}{J. R. Soc. Interface}
  \textbf{\bibinfo{volume}{10}}, \bibinfo{pages}{20120997}
  (\bibinfo{year}{2013}).

\bibitem[{\citenamefont{Milinski et~al.}(2002)\citenamefont{Milinski, Semmann,
  and Krambeck}}]{milinski_n02}
\bibinfo{author}{\bibfnamefont{M.}~\bibnamefont{Milinski}},
  \bibinfo{author}{\bibfnamefont{D.}~\bibnamefont{Semmann}}, \bibnamefont{and}
  \bibinfo{author}{\bibfnamefont{H.-J.} \bibnamefont{Krambeck}},
  \bibinfo{journal}{Nature} \textbf{\bibinfo{volume}{415}},
  \bibinfo{pages}{424} (\bibinfo{year}{2002}).

\bibitem[{\citenamefont{Rezaei and Kirley}(2009)}]{rezaei_ei09}
\bibinfo{author}{\bibfnamefont{G.}~\bibnamefont{Rezaei}} \bibnamefont{and}
  \bibinfo{author}{\bibfnamefont{M.} \bibnamefont{Kirley}},
  \bibinfo{journal}{Evol. Intel.} \textbf{\bibinfo{volume}{2}},
  \bibinfo{pages}{207} (\bibinfo{year}{2009}).

\bibitem[{\citenamefont{Chen et~al.}(2016)\citenamefont{Chen, Chen, and Wang}}]{chen_q_csf16}
\bibinfo{author}{\bibfnamefont{Q.}~\bibnamefont{Chen}},
  \bibinfo{author}{\bibfnamefont{T.}~\bibnamefont{Chen}}, \bibnamefont{and}
  \bibinfo{author}{\bibfnamefont{Y.} \bibnamefont{Wang}},
  \bibinfo{journal}{Chaos Solit. \& Frac.} \textbf{\bibinfo{volume}{91}},
  \bibinfo{pages}{649} (\bibinfo{year}{2016}).

\bibitem[{\citenamefont{Cong et~al.}(2016)\citenamefont{Cong, Li, Wang and Zhao}}]{cong_r_epl16}
\bibinfo{author}{\bibfnamefont{R.}~\bibnamefont{Cong}},
  \bibinfo{author}{\bibfnamefont{K}~\bibfnamefont{Li}},
  \bibinfo{author}{\bibfnamefont{L.}~\bibnamefont{Wang}}, \bibnamefont{and}
  \bibinfo{author}{\bibfnamefont{Q.} \bibnamefont{Zhao}},
  \bibinfo{journal}{EPL} \textbf{\bibinfo{volume}{115}},
  \bibinfo{pages}{38002} (\bibinfo{year}{2016}).

\bibitem[{\citenamefont{Szab{\'o} and T{\H{o}}ke}(1998)}]{szabo_pre98}
\bibinfo{author}{\bibfnamefont{G.}~\bibnamefont{Szab{\'o}}} \bibnamefont{and}
  \bibinfo{author}{\bibfnamefont{C.}~\bibnamefont{T{\H{o}}ke}},
  \bibinfo{journal}{Phys. Rev. E} \textbf{\bibinfo{volume}{58}},
  \bibinfo{pages}{69} (\bibinfo{year}{1998}).

\bibitem[{\citenamefont{Cox and Griffeath}(1986)}]{cox_ap86}
\bibinfo{author}{\bibfnamefont{J.~T.} \bibnamefont{Cox}} \bibnamefont{and}
  \bibinfo{author}{\bibfnamefont{D.}~\bibnamefont{Griffeath}},
  \bibinfo{journal}{Ann. Probab.} \textbf{\bibinfo{volume}{14}},
  \bibinfo{pages}{347} (\bibinfo{year}{1986}).

\bibitem[{\citenamefont{Dornic et~al.}(2001)\citenamefont{Dornic, Chat{\'e},
  Chave, and Hinrichsen}}]{dornic_prl01}
\bibinfo{author}{\bibfnamefont{I.}~\bibnamefont{Dornic}},
  \bibinfo{author}{\bibfnamefont{H.}~\bibnamefont{Chat{\'e}}},
  \bibinfo{author}{\bibfnamefont{J.}~\bibnamefont{Chave}}, \bibnamefont{and}
  \bibinfo{author}{\bibfnamefont{H.}~\bibnamefont{Hinrichsen}},
  \bibinfo{journal}{Phys. Rev. Lett.} \textbf{\bibinfo{volume}{87}},
  \bibinfo{pages}{045701} (\bibinfo{year}{2001}).

\bibitem[{\citenamefont{Szolnoki et~al.}(2009)\citenamefont{Szolnoki, Perc, and
  Szab{\'o}}}]{szolnoki_pre09c}
\bibinfo{author}{\bibfnamefont{A.}~\bibnamefont{Szolnoki}},
  \bibinfo{author}{\bibfnamefont{M.}~\bibnamefont{Perc}}, \bibnamefont{and}
  \bibinfo{author}{\bibfnamefont{G.}~\bibnamefont{Szab{\'o}}},
  \bibinfo{journal}{Phys. Rev. E} \textbf{\bibinfo{volume}{80}},
  \bibinfo{pages}{056109} (\bibinfo{year}{2009}).

\bibitem[{\citenamefont{Szab{\'o} et~al.}(2004)\citenamefont{Szab{\'o},
  Szolnoki, and Izs{\'a}k}}]{szabo_jpa04}
\bibinfo{author}{\bibfnamefont{G.}~\bibnamefont{Szab{\'o}}},
  \bibinfo{author}{\bibfnamefont{A.}~\bibnamefont{Szolnoki}}, \bibnamefont{and}
  \bibinfo{author}{\bibfnamefont{R.}~\bibnamefont{Izs{\'a}k}},
  \bibinfo{journal}{J. Phys. A: Math. Gen.} \textbf{\bibinfo{volume}{37}},
  \bibinfo{pages}{2599} (\bibinfo{year}{2004}).

\bibitem[{\citenamefont{Brandt et~al.}(2003)\citenamefont{Brandt, Hauert, and
  Sigmund}}]{brandt_prsb03}
\bibinfo{author}{\bibfnamefont{H.}~\bibnamefont{Brandt}},
  \bibinfo{author}{\bibfnamefont{C.}~\bibnamefont{Hauert}}, \bibnamefont{and}
  \bibinfo{author}{\bibfnamefont{K.}~\bibnamefont{Sigmund}},
  \bibinfo{journal}{Proc. R. Soc. Lond. B} \textbf{\bibinfo{volume}{270}},
  \bibinfo{pages}{1099} (\bibinfo{year}{2003}).

\bibitem[{\citenamefont{Fu et~al.}(2008)\citenamefont{Fu, Hauert, Nowak, and
  Wang}}]{fu_pre08b}
\bibinfo{author}{\bibfnamefont{F.}~\bibnamefont{Fu}},
  \bibinfo{author}{\bibfnamefont{C.}~\bibnamefont{Hauert}},
  \bibinfo{author}{\bibfnamefont{M.~A.} \bibnamefont{Nowak}}, \bibnamefont{and}
  \bibinfo{author}{\bibfnamefont{L.}~\bibnamefont{Wang}},
  \bibinfo{journal}{Phys. Rev. E} \textbf{\bibinfo{volume}{78}},
  \bibinfo{pages}{026117} (\bibinfo{year}{2008}).

\bibitem[{\citenamefont{Helbing and Yu}(2009)}]{helbing_pnas09}
\bibinfo{author}{\bibfnamefont{D.}~\bibnamefont{Helbing}} \bibnamefont{and}
  \bibinfo{author}{\bibfnamefont{W.}~\bibnamefont{Yu}}, \bibinfo{journal}{Proc.
  Natl. Acad. Sci. USA} \textbf{\bibinfo{volume}{106}}, \bibinfo{pages}{3680}
  (\bibinfo{year}{2009}).

\bibitem[{\citenamefont{Perc}(2011)}]{perc_pre11}
\bibinfo{author}{\bibfnamefont{M.}~\bibnamefont{Perc}}, \bibinfo{journal}{Phys.
  Rev. E} \textbf{\bibinfo{volume}{84}}, \bibinfo{pages}{037102}
  (\bibinfo{year}{2011}).

\bibitem[{\citenamefont{Yang et~al.}(2013)\citenamefont{Yang, Wu, Li, and
  Wang}}]{yang_zh_epjb13}
\bibinfo{author}{\bibfnamefont{Z.}~\bibnamefont{Yang}},
  \bibinfo{author}{\bibfnamefont{T.}~\bibnamefont{Wu}},
  \bibinfo{author}{\bibfnamefont{Z.}~\bibnamefont{Li}}, \bibnamefont{and}
  \bibinfo{author}{\bibfnamefont{L.}~\bibnamefont{Wang}},
  \bibinfo{journal}{Eur. Phys. J. B} \textbf{\bibinfo{volume}{86}},
  \bibinfo{pages}{158} (\bibinfo{year}{2013}).

\bibitem[{\citenamefont{Chen and Szolnoki}(2016)}]{chen_xj_sr16}
\bibinfo{author}{\bibfnamefont{X.}~\bibnamefont{Chen}} \bibnamefont{and}
  \bibinfo{author}{\bibfnamefont{A.}~\bibnamefont{Szolnoki}},
  \bibinfo{journal}{Sci. Rep.} \textbf{\bibinfo{volume}{6}}, \bibinfo{pages}{32802}
  (\bibinfo{year}{2016}).

\bibitem[{\citenamefont{Smaldino}(2013)}]{smaldino_csf13}
\bibinfo{author}{\bibfnamefont{P.~E.} \bibnamefont{Smaldino}},
  \bibinfo{journal}{Chaos, Solit. \& Frac.} \textbf{\bibinfo{volume}{56}},
  \bibinfo{pages}{6} (\bibinfo{year}{2013}).

\bibitem[{\citenamefont{Szolnoki et~al.}(2014)\citenamefont{Szolnoki,
  Antonioni, Tomassini, and Perc}}]{szolnoki_epl14}
\bibinfo{author}{\bibfnamefont{A.}~\bibnamefont{Szolnoki}},
  \bibinfo{author}{\bibfnamefont{A.}~\bibnamefont{Antonioni}},
  \bibinfo{author}{\bibfnamefont{M.}~\bibnamefont{Tomassini}},
  \bibnamefont{and} \bibinfo{author}{\bibfnamefont{M.}~\bibnamefont{Perc}},
  \bibinfo{journal}{EPL} \textbf{\bibinfo{volume}{105}}, \bibinfo{pages}{48001}
  (\bibinfo{year}{2014}).

\bibitem[{\citenamefont{Hofbauer and Sigmund}(1998)}]{hofbauer_98}
\bibinfo{author}{\bibfnamefont{J.} \bibnamefont{Hofbauer}} \bibnamefont{and}
  \bibinfo{author}{\bibfnamefont{K.}~\bibnamefont{Sigmund}},
  \emph{\bibinfo{title}{The Theory of Evolution and Dynamical Systems}} (\bibinfo{publisher}{Cambridge University Press},
  \bibinfo{address}{Cambridge, UK}, \bibinfo{year}{1998}).

\end{thebibliography}
\end{document}